\documentclass[11pt,draftcls,journal,onecolumn]{IEEEtran}
\usepackage{amsmath,amsfonts}
\usepackage{algorithmic}
\usepackage{algorithm}
\usepackage{array}
\usepackage[caption=false,font=normalsize,labelfont=sf,textfont=sf]{subfig}
\usepackage{textcomp}
\usepackage{stfloats}
\usepackage{url}
\usepackage{verbatim}
\usepackage{graphicx}
\usepackage{cite}
\usepackage{standalone}
\usepackage{color}
\usepackage{url}
\usepackage{siunitx}
\usepackage{cite}
\usepackage[capitalise]{cleveref}
\usepackage{fancybox}
        {\begin{Sbox}\begin{minipage}{1.0\textwidth}\parindent1em%
                \noindent\textbf{#1}\noindent}%
        {\end{minipage}\end{Sbox}%
                \begin{figure}[htb!]%
                        \shadowbox{\TheSbox}%
                \end{figure}}
\usepackage{multirow}
\usepackage{booktabs}

\usepackage{pgfplots} % Import of plots from tikzplotlib
\usepackage{tikzsymbols}

\newlength\fheight % Plots from Matlab share the same size
\newlength\fwidth % Plots from Matlab share the same size
\newlength{\figureheight} % Plots from Matlab share the same size
\newlength{\figurewidth} % Plots from Matlab share the same size

\usepackage{tikz}
\pgfplotsset{compat=1.9}

\usetikzlibrary{external}
\usetikzlibrary{arrows}
\usetikzlibrary{patterns}
\usetikzlibrary{backgrounds}
\usetikzlibrary{fit}
\usetikzlibrary{positioning}
\usetikzlibrary{shapes.geometric}
\usetikzlibrary{calc}
\usetikzlibrary{shapes.multipart}
\usetikzlibrary{shapes.misc}
\usetikzlibrary{shapes.symbols}
\usetikzlibrary{patterns.meta}
\usetikzlibrary{fadings}

\pgfdeclarelayer{bg}    % declare background layer
\pgfsetlayers{bg,main}  % set the order of the layers (main is the standard layer)

\tikzset{>=stealth}

\tikzstyle{block}=[
    draw,
    text depth=0pt,
    thick, 
    rectangle,
    text centered,
    minimum height=3.4ex,
    rounded corners=0.3em,
    fill=black!6,
    inner sep=0.6em,
    ] % rectangle

\tikzstyle{dnn}=[]
\tikzstyle{enhBlock}=[]%fill=red]
\tikzstyle{estBlock}=[dashed]%, fill=green]

\tikzstyle{branch}=[{circle,inner sep=0pt,minimum size=0.3em,fill=black}]
\tikzstyle{box}=[rectangle, rounded corners, draw=black, line width=1pt, text width=2cm]

\tikzstyle{arrow}=[{}-{>}, thick]
\tikzstyle{line}=[thick]
\tikzstyle{reverse arrow}=[{<}-{}, thick]

\tikzset{% http://tex.stackexchange.com/a/257632
	do path picture/.style={%
		path picture={%
			\pgfpointdiff{\pgfpointanchor{path picture bounding box}{south west}}%
			{\pgfpointanchor{path picture bounding box}{north east}}%
			\pgfgetlastxy\x\y%
			\tikzset{x=\x/2,y=\y/2}%
			#1
		}
	},
	sin wave/.style={do path picture={    
			\draw [line cap=round] (-3/4,0)
			sin (-3/8,1/2) cos (0,0) sin (3/8,-1/2) cos (3/4,0);
	}},
	cross/.style={draw, circle, do path picture={    
			\draw [line cap=round] (-2/5,-2/5) -- (2/5,2/5) (-2/5,2/5) -- (2/5,-2/5);
	}},
	plus/.style={draw, circle, do path picture={    
			\draw [line cap=round] (-3/5,0) -- (3/5,0) (0,-3/5) -- (0,3/5);
	}},
	mic/.style={inner sep=0pt, do path picture={
			\draw (0,0) circle (0.9);
			\draw [line cap=round] (-0.9, -0.9) -- (-0.9, 0.9);
	}},
	mux/.style={trapezium, draw}
}

\newcommand{\scmtrue}{\ensuremath{\boldsymbol{\Phi}}}
\newcommand{\scmest}{\ensuremath{\hat{\boldsymbol{\Phi}}}}
\newcommand{\instscmest}{\ensuremath{\hat{\boldsymbol{\Psi}}}}

\usepackage{amsmath}
\usepackage{amssymb}
\usepackage{bbm}
\usepackage{siunitx}
\usepackage{trfsigns}

\DeclareMathOperator{\tr}{tr}

\DeclareMathOperator{\ExpOp}{\mathbb{E}}

\newcommand{\vect}[1]{\ensuremath{\boldsymbol{\mathbf{#1}}}}

\newcommand{\HT}{\mathsf{H}}
\newcommand{\T}{\mathsf{T}}

\DeclareUnicodeCharacter{211C}{\Re} %Realteil
\DeclareUnicodeCharacter{2111}{\Im} %Imaginärteil

\definecolor{forestgreen(web)}{rgb}{0.13, 0.55, 0.13}
\colorlet{lightgreen}{green!30}
\colorlet{lightred}{red!30}

\definecolor{diffaddfg}{HTML}{54aeff}
\definecolor{diffaddbg}{HTML}{ddf4ff}
\definecolor{diffdelfg}{HTML}{ffb77c}
\definecolor{diffdelbg}{HTML}{fff1e5}

\definecolor{diffaddfg}{HTML}{007FFF} % Stronger blue
\definecolor{diffdelfg}{HTML}{FF4500} % Stronger orange

\usepackage{ulem}

\makeatletter
\newcommand\hladd{\bgroup\markoverwith{\textcolor{diffaddbg}{\rule[-.5ex]{.1pt}{2.5ex}}}\ULon}
\newcommand\hlsout{\bgroup\markoverwith{\textcolor{diffdelbg}{\rule[-.5ex]{.1pt}{2.5ex}}\llap{\rule[.5ex]{.3pt}{0.4pt}}}\ULon}
\makeatother
\newcommand{\difftwo}[2]{\textcolor{diffdelfg}{{\sout{#1}}}{\color{diffaddfg}{#2}}}
\DeclareRobustCommand{\diff}[2]{#2}
\DeclareRobustCommand{\diffsimple}[2]{#2}
\DeclareRobustCommand{\difftwo}[2]{#2}
\usepackage[acronym, toc, nonumberlist]{glossaries}

\newacronym{AEC}{AEC}{acoustic echo cancellation}
\newacronym{AIR}{AIR}{acoustic impulse response}
\newacronym{AR}{AR}{auto-regressive}
\newacronym{ASR}{ASR}{automatic speech recognition}
\newacronym{ATF}{ATF}{acoustic transfer function}

\newacronym{BLSTM}{BLSTM}{bi-directional long short-term memory}
\newacronym{BLUE}{BLUE}{best linear unbiased estimate}
\newacronym{BSS}{BSS}{blind source separation}

\newacronym{CHiME}{CHiME}{computational hearing in multi-source environments}
\newacronym{CNN}{CNN}{convolutional neural network}
\newacronym{CTF}{CTF}{convolutive transfer function}
\newacronym{CPSD}{CPSD}{cross power spectral density}

\newacronym{DL}{DL}{deep learning}
\newacronym[longplural={directions of arrival}]{DOA}{DoA}{direction of arrival}
\newacronym{DNN}{DNN}{deep neural network}
\newacronym{DNS}{DNS}{deep noise suppresssion}

\newacronym{EM}{EM}{expectation maximization}

\newacronym{GRU}{GRU}{gated recurrent unit}
\newacronym{GSS}{GSS}{guided source separation}

\newacronym{ISCM}{ISCM}{instantaneous spatial covariance matrix}
\newacronym{ICA}{ICA}{independent component analzysis}

\newacronym{LCMV}{LCMV}{linearly constrained minimum variance}

\newacronym{MAXSNR}{max-SNR}{Maximum-SNR}
\newacronym{ML}{ML}{maximum likelihood}
\newacronym{MMSE}{MMSE}{minimum mean squared error}
\newacronym{MTF}{MTF}{multiplicative transfer function}
\newacronym{MVDR}{MVDR}{minimum variance distortionless response}
\newacronym{MWF}{MWF}{multi-channel \diff{wiener}{Wiener} filter}

\newacronym{NN}{NN}{neural network}

\newacronym{PDF}{PDF}{probability density function}
\newacronym{PIT}{PIT}{permutation invariant training}
\newacronym{PSD}{PSD}{power spectral density}

\newacronym{RI}{RI}{real and imaginary}
\newacronym{RIR}{RIR}{room impulse response}
\newacronym{RNN}{RNN}{recurrent neural network}
\newacronym{RTF}{RTF}{relative transfer function}

\newacronym{SAP}{SAP}{source activity probability}
\newacronym[longplural={spatial covariance matrices}]{SCM}{SCM}{spatial covariance matrix}
\newacronym{SDW}{SDW}{speech distortion weighted}
\newacronym{SNR}{SNR}{signal-to-noise power ratio}
\newacronym{SPP}{SPP}{speech presence probability}
\newacronym{STFT}{STFT}{short time Fourier transform}
\newacronym{TDOA}{TDoA}{time difference of arrival}
\newacronym{tf}{tf}{time-frequency}
\newacronym{TSVAD}{TS-VAD}{target speaker voice activity detection}

\newacronym{VAD}{VAD}{voice activity detection}

\newacronym{WPD}{WPD}{weighted power minimization distortionless response}
\newacronym{WPE}{WPE}{weighted prediction error}

\begin{document}
\providecommand{\myTitle}{Microphone Array Signal Processing and Deep Learning for Speech Enhancement}

\title{\myTitle}

\author{\IEEEauthorblockN{Reinhold Haeb-Umbach$^*$, Tomohiro Nakatani$^\dagger$, Marc Delcroix$^\dagger$, Christoph Boeddeker$^*$, Tsubasa Ochiai$^\dagger$\\}
\IEEEauthorblockA{
\textit{$^*$Paderborn University, Germany; $^\dagger$NTT Corporation, Japan}
}
}

%\author{Reinhold Haeb-Umbach,
%	Tomohiro Nakatani,
%	Marc Delcroix,
%	Christoph Boeddeker,\\
%	Tsubasa Ochiai}% <-this % stops a space

\markboth{DRAFT -- \myTitle}%
{Shell \MakeLowercase{\textit{et al.}}: A Sample Article Using IEEEtran.cls for IEEE Journals}

\IEEEpubid{0000--0000/00\$00.00~\copyright~2021 IEEE}

\maketitle

\begin{abstract}

Multi-channel acoustic signal processing is a well-established and powerful tool to exploit the spatial diversity between a target signal and\diff{}{  non-target or} noise sources for signal enhancement.
However, the textbook solutions for optimal data-dependent spatial filtering rest on the knowledge of second-order statistical moments of the signals, which have traditionally been difficult to acquire.
In this contribution, we compare model-based, purely data-driven, and hybrid approaches to parameter estimation and filtering, where the latter tries to combine the benefits of model-based signal processing and data-driven deep learning to overcome their individual deficiencies.
We illustrate the underlying design principles with examples from noise reduction, source separation, and dereverberation.
%While today's data-driven solutions give excellent enhancement performance, the blending with model-based components can add to the robustness of the overall system.
%We conclude with a critical discussion of the current trend towards fully data-driven (all-neural) non-linear spatial filtering.

%\inred{This is the one from the white paper -- has not yet been adapted} %Multi-channel acoustic signal processing is a well-established and powerful tool to exploit the spatial diversity between a target signal and noise sources for signal enhancement.
% However, the textbook solutions for optimal data-dependent spatial filtering rest on the knowledge of second-order statistical moments of the signals, which have traditionally been difficult to acquire.
% This is, where neural networks excel: they can effectively learn a nonlinear mapping from input to target values and thus provide the required parameter estimates. This eminently fruitful collaboration of deep learning and statistical signal processing has developed since its first encounter in neural network supported acoustic beamforming into a very active research area, with many new algorithms and applications.
% This tutorial will give an overview of this exciting development.
 
\end{abstract}

\begin{IEEEkeywords}
Microphone array processing, speech enhancement, deep learning, source separation
\end{IEEEkeywords}

\section{Introduction}

Speech enhancement has been a research topic for many years.
It is concerned with removing distortions, such as ambient noise, from a degraded speech signal.
Its various applications fall into two classes: improving human-to-human communication and improving human-to-machine communication, i.e., \gls{ASR}.
In the first case, where the human is the consumer of the enhanced signal, the focus is on enhancing the quality, and ideally also the intelligibility, of the signal captured by the microphone(s).
Examples of applications are telecommunications, hearing aids, media archive enhancement, etc.\diffsimple{}{ \cite{graetzer2021clarity,somasundaram2023projectAria,yoshioka2022vararray,Haeb-Umbach_2019}.}
If the enhancement stage is followed by an \gls{ASR} system, the goal is to improve the recognition accuracy of the recognizer.

There are many ways in which a speech signal can be distorted.
It can be degraded by the superposition of other acoustic signals, such as by environmental noise or competing speakers.
The multi-path propagation from the speech source to the microphones, known as reverberation, leads to degraded intelligibility and \gls{ASR} performance.
And there are many other distortions, e.g., line or acoustic echoes, signal clipping, etc.
Consequently, the field of speech enhancement is large, and a plethora of techniques have been proposed to clean up a speech signal.

Those techniques can be classified according to the enhancement task to be solved.
In this contribution, we concentrate on noise reduction, source separation, and dereverberation.
Another classification is into methods using a single microphone (single-channel) or multiple microphones (multi-channel).
Here, we are concerned with multi-channel approaches, assuming that the signals are captured by a compact microphone array.
Finally, and this is the focus of this study, methods can be categorized into model-based and data-driven methods, as well as hybrid methods that combine the two.

Research in speech enhancement has a long tradition, with the spectral subtraction method by Boll \cite{boll1979suppression} being an early hallmark.
This heuristic approach was soon complemented by statistically motivated methods \cite{ephraim1984speech}.
For years, research was dominated by such model-based approaches, which use statistical signal processing methods derived from a physical model of the degradation and a probabilistic model of the involved signals.
With the advent of deep learning, those model-based techniques have been challenged by data-driven approaches that learn the enhancement operation from data.
This is typically done by training a \gls{DNN} in a supervised manner with the observed degraded signal at its input and the desired signal or a quantity derived from it as the training target.
A third class of techniques, hybrid approaches, blend model-based with data-driven methods.
They aim to combine the benefits of both approaches while overcoming their individual shortcomings.

This paper is not about giving a comprehensive overview of the various approaches to speech enhancement.
Its purpose is rather to highlight the advantages and shortcomings of the above classes of enhancement techniques mentioned in the last section and discuss those with a few concrete examples from the field of multi-channel enhancement.
We look into noise reduction, dereverberation, and source separation, for which we can juxtapose the three classes of enhancement methods, as outlined in Table~\ref{table:examples}.
\diff{}{Although many presented methods have more applications than being used for speech, e.g., for music signals, we focus here on speech for brevity.}

\begin{table}[t]\centering
\caption{Examples of model-based, data-driven, and hybrid speech enhancement approaches for Noise Reduction (NR), Dereverberation (DR), and Source Separation (SS).
The related \diff{chapters}{sections} in this paper \diff{}{and pros and cons of each approach} are also indicated.}\label{table:examples}
% \begin{tabular}{lllllll}\toprule
%   \multicolumn{2}{l}{Model-based} & \multicolumn{2}{l}{Data-driven} & \multicolumn{2}{l}{Hybrid}\\\midrule
%  \begin{minipage}{1cm}
%    \ref{sec:model_beamforming}\\\\\\
%    \ref{sec:model_beamforming}\\
%    \ref{sec:model_dereverberation}\\\\\\\\
%  \end{minipage} & \hspace{-3.5mm}\begin{minipage}{3.5cm}\flushleft
%    Beamforming (BF) with EM-based mask estimation for NR and SS\\
%    Blind SS (BSS) for SS\\
%    Linear prediction (LP), Kalman filter for DR\\~\\~\\~\\
%  \end{minipage} & \begin{minipage}{1cm}
%    \ref{sec:data_denoising}\\\\\\
%    \ref{sec:data_denoising}\\\\
%    \ref{sec:beamforming_case_study}\\\\\\
%  \end{minipage} & \hspace{-3.5mm}\begin{minipage}{4cm}\flushleft
%    Spectral mapping and masking by \gls{DNN} for NR and DR\\
%    Permutation Invariant Training (PIT) for SS\\
%    Trend to purely data-driven approach\\~\\~\\
%  \end{minipage} &\begin{minipage}{1cm}
%    \ref{sec:joint_model_data_driven_enhancement}\\\\
%    \ref{sec:joint_model_data_driven_parameter}\\~\\
%    \ref{sec:joint_model_data_driven_parameter}\\~\\
%    \ref{sec:modeldeficiencies}\\~\\
%    \ref{sec:end-to-end}
%  \end{minipage} & \hspace{-3.5mm}\begin{minipage}{4.5cm}\flushleft
%    BF with \gls{DNN}-based mask estimation for NR and SS \\
%    BF with EM and \gls{DNN}-based mask estimation for NR and SS \\
%    \gls{DNN}-guided SS/LP for NR, DR and SS\\
%    Spectral mapping by \gls{DNN}-BF-\gls{DNN}  for NR, DR, and SS \\
%    End-to-end optimization\\
%  \end{minipage}\\\bottomrule
% \end{tabular}
\begin{tabular}{lll}\toprule
  Model-based & Data-driven & Hybrid\\
  \midrule
  \begin{tabular}[t]{@{}l@{\hskip 0.7em}>{\raggedright\arraybackslash}p{3.5cm}@{}} % "@{}" omits side padding
       \ref{sec:model_beamforming} & Beamforming (BF) with EM-based mask estimation for NR and SS \\
       \ref{sec:model_beamforming} & Blind SS (BSS) for SS \\
       \ref{sec:model_dereverberation} & Linear prediction (LP), Kalman filter for DR
  \end{tabular}
  &
  \begin{tabular}[t]{@{}l@{\hskip 0.7em}>{\raggedright\arraybackslash}p{3.5cm}@{}}
      \ref{sec:data_denoising} & Spectral mapping and masking by \gls{DNN} for NR and DR \\
      \ref{sec:data_denoising} & Permutation Invariant Training (PIT) for SS \\
      \ref{sec:beamforming_case_study} & Trend to purely data-driven approach
  \end{tabular}
  &
  \begin{tabular}[t]{@{}l@{\hskip 0.7em}>{\raggedright\arraybackslash}p{5.2cm}@{}}
    \ref{sec:joint_model_data_driven_enhancement} & BF with \gls{DNN}-based mask estimation for NR and SS \\
    \ref{sec:joint_model_data_driven_parameter} & BF with EM and \gls{DNN}-based mask estimation for NR and SS \\
    \ref{sec:joint_model_data_driven_parameter} & \gls{DNN}-guided SS/LP for NR, DR and SS \\
    \ref{sec:modeldeficiencies} & Spectral mapping by \gls{DNN}-BF-\gls{DNN}  for NR, DR, and SS \\
    \ref{sec:end-to-end} & End-to-end optimization
  \end{tabular}\\
  \midrule
  \begin{tabular}[t]{@{}l@{\hskip 0.7em}>{\raggedright\arraybackslash}p{3.5cm}@{}} % "@{}" omits side padding
       \diff{}{Pros} & \diff{}{High adaptability, explainability} \\
       \diff{}{Cons} & \diff{}{Restricted by model assumptions}
  \end{tabular}
  &
  \begin{tabular}[t]{@{}l@{\hskip 0.7em}>{\raggedright\arraybackslash}p{3.5cm}@{}}
      \diff{}{Pros} & \diff{}{No model assumptions, high performance} \\
      \diff{}{Cons} & \diff{}{Requires a large amount of training data, sensitivity to train-test mismatch}
  \end{tabular}
  &
  \begin{tabular}[t]{@{}l@{\hskip 0.7em}>{\raggedright\arraybackslash}p{5.2cm}@{}}
    \diff{}{Pros} & \diff{}{Fewer model assumptions, high performance with high adaptability} \\
    \diff{}{Cons} & \diff{}{Requires a large amont of training data, adaptability is partly limited by the data-driven approach}
  \end{tabular}
  \\
 % \end{minipage} & \begin{minipage}{1cm}
 %   \\\\\\
 %   \\\\
 %   \\\\\\
 % \end{minipage} & \hspace{-3.5mm}\begin{minipage}{4cm}\flushleft
 %   \\
 %   \\
 %   \\~\\~\\
 % \end{minipage} &\begin{minipage}{1cm}
 % \end{minipage} & \hspace{-3.5mm}\begin{minipage}{4.5cm}\flushleft
 %    \\
 %    \\
 %   \\
 %    \\
 %   \\
 % \end{minipage}\\
 \bottomrule
\end{tabular}
\end{table}
We will start with the presentation of commonly used signal models.
The models have a decreasing degree of complexity.
We continue by describing the properties of model-based and data-driven enhancement approaches.
We then put an emphasis on hybrid methods, for which we offer a taxonomy.
In recent research, we are observing a trend towards data-driven methods, replacing more and more components of the enhancement operation with deep learning-based alternatives.
We exemplify this trend with the example of acoustic beamforming and finish with a critical discussion of this trend.

\section{Signal models}
\label{sec:signal_models}
%%%%%%%%%%%%%%%%%%%%%%%%%%%%%%%%%%%%%%%%%%%

\begin{figure}
    \centering
    \input{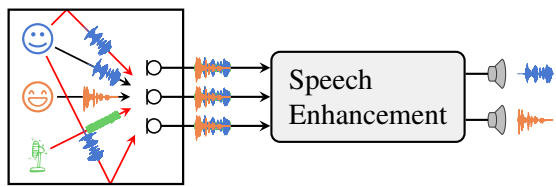}
    \caption{Illustration of the considered scenario: A reverberant enclosure with desired and undesired acoustic sources.
    The signal is captured by a microphone array and processed by a speech enhancement system to output one or more cleaned-up signals.}
    \label{fig:far_field_scenario}
\end{figure}

At the outset of any speech enhancement method is a physical and/or statistical model that describes the properties of the desired signal, the distortion, and how the distortion impacts the desired signal.
As shown in \cref{fig:far_field_scenario}, we consider a scenario where the microphone signal is distant from the speakers, i.e., a far-field scenario, as opposed to a close-talking scenario.

The signals at the microphones are modeled as a linear superposition of the convolution of in general $K$ source signals  $\tilde{s}_{k,\ell}, k=1, \ldots , K$ (including, of course, the case of a single source, i.e., $K=1$)  with their respective \glspl{AIR}  \difftwo{$\tilde{\vect{h}}_{k,\ell,\tilde{\ell}}$}{$\tilde{\vect{h}}_{k,\tilde{\ell}}$} from a source $k$ to the $D$ microphones and an additive distortion $\tilde{\vect{n}}_\ell$
%\begin{align}
%	\label{eq:time_domain_convolution}
%	\tilde{\vect{y}}_\ell = \sum_{k=1}^K\sum_{\tilde{\ell}=0}^{\tilde{L}-1}\tilde{\vect{h}}_{k,\ell,\tilde{\ell}}\tilde{s}_{k,\ell-\tilde{\ell}} + \tilde{\vect{n}}_\ell \in \mathbb{R}^{D},
%\end{align}
\begin{align}
	\label{eq:time_domain_convolution}
	\tilde{\vect{y}}_\ell = \sum_{k=1}^K\sum_{\tilde{\ell}=0}^{\tilde{L}-1}\difftwo{}{\tilde{\vect{h}}_{k,\tilde{\ell}}}\tilde{s}_{k,\ell-\tilde{\ell}} + \tilde{\vect{n}}_\ell \in \mathbb{R}^{D},
\end{align}
where $\ell$  and $\tilde{\ell}$ are the discrete time and the \gls{AIR}'s tap indices, respectively.
Here, $\tilde{\vect{y}}_\ell$, \difftwo{}{$\tilde{\vect{h}}_{k,\tilde{\ell}}$} and $\tilde{\vect{n}}_\ell$ are $D$-dimensional vectors of the time-domain signals at the microphones, the \gls{AIR}, and the noise.
\diff{Note that one has to assume in general that the  \glspl{AIR}, $\tilde{\vect{h}}_{k,\ell,\tilde{\ell}}$, are time-varying.}{}
\diff{They}{The \glspl{AIR}} represent the propagation of the speech signal from the source to the microphones.
This is in general a multi-path propagation, seen in the equation by the convolution of the source signal with the \gls{AIR}, resulting in a reverberated signal at the sensors.
\difftwo{Note that one has to assume in general that the  \glspl{AIR}, $\tilde{\vect{h}}_{k,\ell,\tilde{\ell}}$, are time-varying to model, e.g., speaker movements or changes in the environment.}{Note that we assumed a time-invariant \gls{AIR} for simplicity.}
% \diff{}{If a source, a microphone or an object in the room moves, }

The speech signals are either treated as deterministic variables or as random processes, where Gaussian and supergaussian \glspl{PDF} are typical assumptions.
The noise $\tilde{\vect{n}}_\ell$ is usually modeled as a Gaussian random process, which is independent of the speech.

With \cref{eq:time_domain_convolution}, the different enhancement tasks can be explained: 
\begin{itemize}
    \item Noise reduction refers to removing $\tilde{\vect{n}}$ from $\tilde{\vect{y}}$.
    If this is done by exploiting the multi-channel recording instead of only a single channel, it is called beamforming or spatial filtering.
    %In this contribution we use ``spatial filtering'' as the general term and call it ``beamforming'' only if it is done with a linear filter.
    \item  Source separation is concerned with isolating the individual source signals $\tilde{s}_k, k=1,\ldots , K$\diff{}{ or the reverberated source signals at the microphones}, from the mixture $\tilde{\vect{y}}$.
    If we are only interested in one of the sources, say \diff{$\tilde{s}_1$}{$k=1$}, we call it source extraction.
    In that case, the remaining source signals $\tilde{s}_k$, $k =2, \ldots , K$ are viewed as additive distortions.
    \item Dereverberation is about reducing the reverberation components in the \gls{AIR} \difftwo{}{$\tilde{\vect{h}}_{k,\tilde{\ell}}$}, such that the estimated source signal is not corrupted by delayed versions of it.
\end{itemize}

While \cref{eq:time_domain_convolution}  can be considered a faithful model, it is often too complex to be used in practice for deriving enhancement algorithms.
According to the well-known saying of the statistician George Box that ``Essentially all models are wrong, but some are useful'' \cite{box1976science}, virtually all models are only a more or less exact approximation to the reality, and it is important to know the assumptions a model makes.

As a first step towards tractability, the signals are represented in the \gls{STFT} domain.
Since speech is a broadband signal, its processing is simplified if done in the \gls{STFT} domain, where the time-domain convolution can be approximated by a convolution over the frame index $t$ employing a much shorter \gls{ATF} $\vect{h}$ (so-called \gls{CTF} approximation \cite{avargel2007system}):
\begin{align}
	\label{eq:ctf}
	\vect{y}_{t,f} &= \sum_{k=1}^K \underbrace{\sum_{\tau = 0}^{L-1}\vect{h}_{k,\tau,f}s_{k,t-\tau,f}}_{\vect{x}_{k,t,f}} + \vect{n}_{t,f}  \\
    \label{eq:ctf_2}
    &= \sum_{k=1}^K \left(\sum_{\tau = 0}^{\Delta-1}\vect{h}_{k,\tau,f}s_{k,t-\tau,f} + \sum_{\tau = \Delta}^{L-1}\vect{h}_{k,\tau,f}s_{k,t-\tau,f}\right)  + \vect{n}_{t,f}  \\
    \label{eq:ctf_3}
    &= \sum_{k=1}^K \left( \vect{x}_{k,t,f}^{\text{(early)}} + \vect{x}_{k,t,f}^{\text{(late)}}\right) + \vect{n}_{t,f} \; \in \mathbb{C}^D.
\end{align}
Here, $t$ and $f$ are the time frame and frequency bin indices, and $s_{k,t,f}$, $\vect{y}_{t,f}$ $\vect{n}_{t,f}$ are the \gls{STFT} of the $k$-th source signal and the \gls{STFT} vectors of the microphone signal and the noise, respectively\difftwo{.}{, while $\vect{h}_{k,t,f}$ is the vector of \glspl{ATF}.}
\diff{}{The time lag $\Delta$ separates the direct signal and early reflections from the late reverberation. Dereverberation is typically  concerned with removing the late reverberation only.}
\difftwo{We further assumed that the \gls{AIR} is time-invariant, and so is the \gls{ATF} vector $\vect{h}_{k,t,f}$.}{}
\Cref{eq:ctf} is an approximation to \cref{eq:time_domain_convolution}, because the cross-band filters, which are used for canceling the aliasing effects caused by subsampling from sampling rate to frame rate, have been neglected%.
%This can be justified if the signals vary slowly over time
, see \cite{avargel2007system} for a discussion.
We further introduced the notation $\vect{x}_{k,t,f}$ for the spatial image of the source signal $s_{k,t,f}$ at the microphone array, which can be divided in $\vect{x}_{k,t,f}^{\text{(early)}}$, the direct signal plus early reflections, and $\vect{x}_{k,t,f}^{\text{(late)}}$, the late reverberation.

The model can be further simplified by approximating the convolution over the frame index by a multiplication  (so-called \gls{MTF} or narrowband  approximation \cite{avargel2007system, gannot2017consolidated}):
\begin{align}
	\label{eq:mtf}
	\vect{y}_{t,f} &= \sum_{k=1}^K \vect{h}_{k,f}\vect{s}_{k,t,f} + \vect{n}_{t,f} 
	= \sum_{k=1}^K \vect{x}_{k,t,f} + \vect{n}_{t,f}.
\end{align}
This simplification is justified if the \gls{STFT} frame length is (significantly) larger than the length of the \gls{AIR}, i.e., in low reverberant situations.

The model can be even further simplified by assuming an anechoic environment, i.e., neglecting reverberation completely.
Employing the far-field approximation, which says that the distance of the source to the microphones is much larger than the inter-microphone distance, the attenuation through signal propagation is the same for all elements and can be neglected for the model.
Then, the vector of \glspl{ATF} consists of complex exponentials representing delays caused by the different propagation path lengths from the sources to the sensors of the microphone array, which depend on the \glspl{DOA} of the sources:
\begin{align}
	\label{eq:anechoic}
	\vect{h}_{k,f} = [\e^{-j2\pi \tilde{f}\tau_{k,1}}, \e^{-j2\pi \tilde{f}\tau_{k,2}}, \ldots , \e^{-j2\pi \tilde{f}\tau_{k,D}}]^\T.
\end{align}
Here, $\tau_{k,d}$ is the propagation delay from the $k$-th source to the $d$-th microphone, which depends on the \gls{DOA}, and $\tilde{f} = f\cdot (f_s/F)$ is the frequency in Hertz ($f_s$: sampling frequency, $F$: DFT size).

Another simplification of \eqref{eq:mtf} is to neglect the noise and concentrate on reverberation: %In the case of a single source ($K=1$) this is
%\begin{align}
%	\label{eq:reverb_only}
%	\vect{y}_{t,f} =  \sum_{\tau = 0}^{L-1}\vect{h}_{\tau,f}s_{t-\tau,f} = \vect{x}_{k,t,f}^{\text{(early)}} + \vect{x}_{k,t,f}^{\text{(late)}}.
%\end{align}
\begin{align}
	\label{eq:reverb_only}
	\vect{y}_{t,f} =  \sum_{k=1}^K\sum_{\tau = 0}^{L-1}\vect{h}_{\tau,f}s_{t-\tau,f} = \sum_{k=1}^K\vect{x}_{k,t,f}^{\text{(early)}} + \sum_{k=1}^K\vect{x}_{k,t,f}^{\text{(late)}}.
\end{align}

Eqs.~\eqref{eq:time_domain_convolution}~-~\eqref{eq:reverb_only} are physical models of different degrees of complexity.
A model of decreasing complexity goes hand in hand with the model having an increasing degree of approximations.
Obviously, the desire for a simple, tractable model collides here with the desire of high faithfulness.

%%%%%%%%%%%%%%%%%%%%%%%%%%%%%%%%%%%%%%%%%%%%%
\section{Model-based and data-driven approaches to enhancement}
\label{sec:model_data_hybrid}
%%%%%%%%%%%%%%%%%%%%%%%%%%%%%%%%%%%%%%%%%%%%

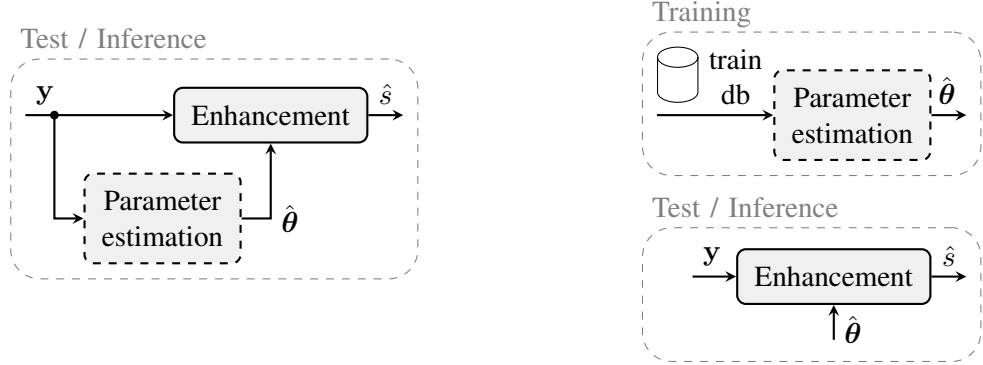
\begin{figure}
	\begin{center}
	{
\linespread{1.}

\begin{tikzpicture}
    \node[block, enhBlock] (enh) at (0, 0) {Enhancement};

    \node[block,estBlock,align=center,anchor=north east] (est) at ($(enh.south)+(-1em,-1em)$) {Parameter\\estimation};

    \draw [arrow] (est) -| node[right] {$\hat{\vect{\theta}}$} (enh);

    \draw [arrow] (enh.east) -- +(1.2em,0) node[above left] (shat) {$\hat{s}$};

    \draw[arrow] (est.west |- enh) +(-2em,0) node[above right] (y) {$\vect{y}$} -- (enh);

    \draw[arrow] (est.west |- enh) +(-1em,0) node[branch] {} |- (est);

    \node[fit=(enh)(est)(shat)(y), inner sep=0.5em, draw, dashed, gray, rounded corners=1em, shift={(0,0em)}] (textbox) {};
    \node[text=gray, above right, anchor=south west, rotate=0] at (textbox.north west) {Test / Inference};

    \begin{scope}[xshift=20em]
        \node[block,estBlock,align=center, dnn] (est) at (0, 0) {Parameter\\estimation};
        \draw[arrow] (est.east) -- +(1.2em,0) node[above left] (labeltheta) {$\hat{\vect{\theta}}$};

        \draw[arrow] (est.west) +(-4em,0) -- +(0,0) node [above left, align=center] (dbtext) {train\\db};

        \node[draw,cylinder,shape border rotate=90,aspect=0.5,anchor=east] (db) at ($(dbtext.north west)!0.6!(dbtext.south west)$) {\phantom{w}};

        \node[block,enhBlock,anchor=north east, dnn] (enh) at ($(est.south east)+(0,-3.0em)$) {Enhancement};
        \draw[arrow] (enh.east) -- +(1.2em,0) node[above left] (labelshat) {$\hat{s}$};
        \draw[arrow] (enh.south) +(0,-1.2em)  -- +(0,0) node[below right] (labeltheta2) {$\hat{\vect{\theta}}$};
        \draw[arrow] (enh.west) +(-1.5em,0) node[above right] {$\vect{y}$} -- +(0,0);

        \node[fit=(db.north)(db.west)(est.south)(est.east)(labeltheta), inner sep=0.5em, draw, dashed, gray, rounded corners=1em, shift={(0,0em)}] (trainbox) {};
        \node[text=gray, above right, anchor=south west, rotate=0] at (trainbox.north west) {\vphantom{T}\smash{Training}};

        \coordinate (tmp) at ($(labeltheta2.south)+(0,0.25em)$);
        \coordinate (tmp2) at ($(labelshat.north east)-(0,0.25em)$);
        \node[fit=(db.west|-enh)(enh)(tmp)(tmp2), inner sep=0.5em, draw, dashed, gray, rounded corners=1em, shift={(0,0em)}] (textbox) {};
        \node[text=gray, above right, anchor=south west, rotate=0] at (textbox.north west) {Test / Inference};
    
    \end{scope}
    
\end{tikzpicture}
}
	\caption{A speech enhancement system consists of two operations, the estimation of the parameters $\boldsymbol{\theta}$ of the enhancement operation (dashed boxes) and the actual enhancement (solid boxes).
    Left: the parameters are estimated from the same signal that has to be enhanced.
    This is the typical scenario in model-based approaches.
    Right: Parameters are estimated in a training stage and then used in the enhancement operation on the test data.
    This is typical for data-driven approaches.}
	\label{fig:enhancement_system}
	\end{center}
\end{figure}

Speech enhancement, as many other signal processing tasks, consists of two subtasks: the estimation of the parameters of the enhancement algorithm and the actual enhancement operation, see \cref{fig:enhancement_system}.

For the first task, there are two options: either the parameters are estimated from the very same signal that is to be enhanced, or the parameters are estimated in a separate training stage.
Of course, there is also the option that the parameters estimated in the training stage are adapted during the enhancement of the test data.
We will come back to this option later on.

%For the actual enhancement operation, one can discern model-based or signal processing based, and data-driven or neural network based methods.
% There exist also hybrid approaches that are a mix of the two.

Model-based parameter estimation and enhancement is done as illustrated in the left figure, while
data-driven estimation and enhancement take the route illustrated in the right figure.
Hybrid methods blend model-based with data-driven approaches, where this blending can be done either in the parameter estimation, in the enhancement stage, or in both.

%%%%%%%%%%%%%%%%%%%%%%%%%%%%%%%%%%%%%%%%%%%%%%%%%%
\subsection{Model-based parameter estimation and enhancement}
\label{sec:modelbased}
%%%%%%%%%%%%%%%%%%%%%%%%%%%%%%%%%%%%%%%%%%%%

For years, many sophisticated model-based enhancement techniques have been developed.
Given a signal model, estimation theory is employed to derive statistically optimal estimators.
Being able to prove the optimality of an estimator is an important property, which most data-driven techniques lack.
Note, however, that optimality rests upon certain assumptions, foremostly the validity of the model and the knowledge of its parameters.
Below, we introduce representative examples of model-based approaches for denoising, source separation, and dereverberation.

%%%%%%%%%%%%%%%%%%%%%%%%%%%%%%%%%%%%%%%%%%%%%%%%%%
\subsubsection{Acoustic beamforming for denoising and source separation}
\label{sec:model_beamforming}
%%%%%%%%%%%%%%%%%%%%%%%%%%%%%%%%%%%%%%%%%%%%

Beamforming exploits the spatial diversity of the sources and points a beam  towards the source of interest and enhances that source signal, while suppressing signals with a different spatial signature.
It performs denoising and/or source separation/extraction depending on whether the signals to be suppressed include noise and/or interfering source signals.
For this problem, optimal solutions have been derived
%\difftwo{, see the box entitled ``Statistically Optimum Beamforming'' on page~\pageref{sidebar:statistically_optimum_beamforming}.}{\cite{Gannot_2017}.}
\cite{gannot2017consolidated}.
\difftwo{}{A common choice for the beamformer is the \gls{MVDR} beamformer. Its coefficients can be computed as follows \cite{souden2009optimal}
\begin{align}
	\label{eq:mvdr}
	\vect{w}_{f}^{\text{MVDR}} &=  \frac{\left(\scmtrue_{\vect{n},f}\right)^{-1}\scmtrue_{\vect{x},f}}{ \tr\left\{\left(\scmtrue_{\vect{n},f}\right)^{-1}\scmtrue_{\vect{x}_f}\right\}} \vect{u}_1
\end{align}
where $\vect{u}_1$ is an all-zero vector with a one at the position of the reference microphone, and $\tr\{\cdot\}$ is the trace operator.
In these equations, $\scmtrue_{\vect{x},f}= \ExpOp[\vect{x}_{t,f}\vect{x}_{t,f}^\HT]$ and $\scmtrue_{\vect{n},f}=\ExpOp[\vect{n}_{t,f}\vect{n}_{t,f}^\HT]$ are the \glspl{SCM} of the speech and the noise part of the microphone signals, respectively.}

The key problem in beamforming is to estimate the beamformer coefficients $\vect{w}_f$ that allow the extraction of the desired signal.
This boils down to estimating the \glspl{SCM}  $\scmtrue_{\vect{x},f}$ and $\scmtrue_{\vect{n},f}$ of the desired signal and the noise, respectively.

\begin{figure}
    \centering
    % \documentclass[preview]{standalone}

% \input{style/tikz_header}

% \begin{document}

% \setlength{\figurewidth}{20em}
% \setlength{\figureheight}{20em}

% \input{figures/example_sep_observation}
% \input{figures/example_sep_mask1}
% \input{figures/example_sep_mask2}
% \input{figures/example_sep_masknoise}

\linespread{1.}

\definecolor{darkslategray38}{RGB}{38,38,38}
\definecolor{lightgray204}{RGB}{204,204,204}
\pgfplotsset{
    example_sep_common_axis_opts/.style={
        axis line style={lightgray204},
        width=13em,
        height=12em,
        % xticklabel distance=1cm,
        tick align=outside,
        x grid style={lightgray204},
        xlabel style={align=center},
        xmajorticks=true,
        xmin=-0.5, xmax=265.5,
        ticklabel style={font=\tiny},
        label style={font=\small},
        major tick length=0.1em,
        ytick distance={100},
        xminorgrids,
        % xtick style={color=darkslategray38},
        ticklabel style={inner sep=0,outer sep=0.1em},
        title style={inner sep=0pt,outer sep=0},
        y grid style={lightgray204},
        ymajorticks=false,
        ymin=-0.5, ymax=512.5,
        % yminorgrids,
        % ytick style={color=darkslategray38}
    },
    plot graphics/example_sep_common_plot_opts/.style={
        includegraphics cmd=\pgfimage,xmin=-0.5, xmax=265.5, ymin=-0.5, ymax=512.5
    }
}

\begin{tikzpicture}
\begin{axis}[
    example_sep_common_axis_opts,
    ymajorticks=true,
    ylabel=\textcolor{darkslategray38}{Frequency bin index f},
    xlabel={Time frame index t},
    title={$\vect{y}_{t, f}$},
    point meta max=2.73760203489008,
    point meta min=-57.2623979651099,
    colorbar,
    colorbar/width=0.5em,
    % colorbar style={ylabel={Energy / dB}},
    colormap/viridis,
    % colorbar horizontal,    
    colorbar style={
        yticklabel={$\pgfmathprintnumber{\tick}$ dB},
        yticklabel style={rotate=270},
        at={(1.04,0)},anchor=south west},
]
\addplot graphics [example_sep_common_plot_opts] {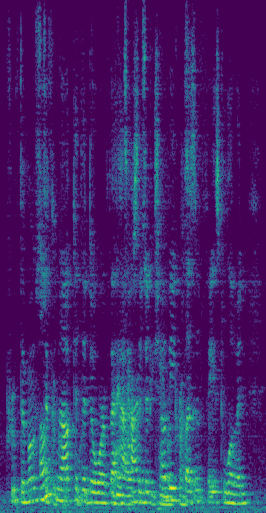};
\end{axis}
\end{tikzpicture}
\begin{tikzpicture}
\begin{axis}[
    example_sep_common_axis_opts,
    xlabel={Time frame index t},
    title={$\vect{m}_{1, t, f}$},
]
\addplot graphics [example_sep_common_plot_opts] 
% {figures/example_sep_mask1-000.png};
{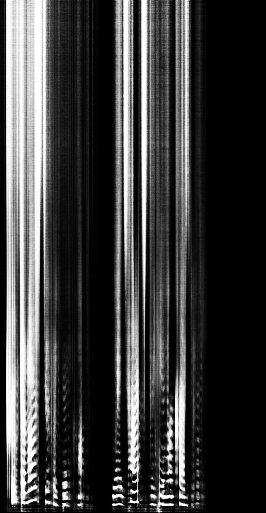};
\end{axis}
\end{tikzpicture}
\begin{tikzpicture}
\begin{axis}[
    example_sep_common_axis_opts,
    xlabel={Time frame index t},
    title={$\vect{m}_{2, t, f}$},
]
\addplot graphics [example_sep_common_plot_opts] 
% {figures/example_sep_mask2-000.png};
{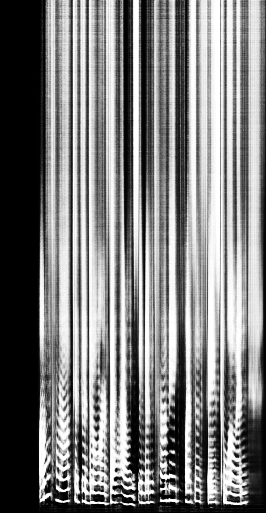};
\end{axis}
\end{tikzpicture}
\begin{tikzpicture}
\begin{axis}[
    example_sep_common_axis_opts,
    xlabel={Time frame index t},
    title={$\vect{m}_{0, t, f}$},
    point meta max=1,
    point meta min=0,
    colorbar,
    colorbar/width=0.5em,
    % colorbar style={ylabel={Energy / dB}},
    % colormap/viridis,
    colormap/blackwhite,
    % colorbar horizontal,    
    colorbar style={
        yticklabel={$\pgfmathprintnumber{\tick}$},
        yticklabel style={rotate=270},
        at={(1.04,0)},anchor=south west},
]
\addplot graphics [example_sep_common_plot_opts] 
% {figures/example_sep_masknoise-000.png};
{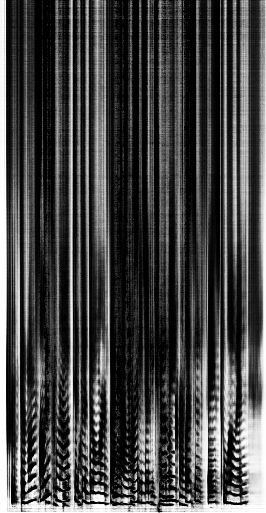};
\end{axis}
\end{tikzpicture}

% \end{document}
    \caption{A 4.25 seconds long segment of speech from the LibriCSS corpus \cite{chen2020continuous}.
    From left (1) to right (4): (1) Observation, (2) mask of first active speaker, (3) mask of second active speaker, (4) the noise mask.
    Masks were estimated by a \gls{DNN}.}
    % computed as one minus the speaker masks
    \label{fig:masks}
    % overlap_ratio_40.0_sil0.1_1.0_session0_actual39.5, start: 32.2, stop: 36.45
\end{figure}

One popular approach to estimating these quantities is an indirect one, which first estimates a signal activity parameter, from which the variance parameters can then be computed.
This indirect method is based on the sparseness property and the w-disjoint orthogonality of speech in the \gls{STFT} domain \cite{yilmaz2004blind}, which say that a \gls{tf}-bin is either dominated by speech or by noise, and in case of multiple speakers, the individual speakers are dominant in disjoint sets of \gls{tf}-bins.
This can be expressed by introducing a latent one-hot vector $\vect{z}_{t,f}\in \{0,1\}^{K+1}$, where $z_{k,t,f}=1$ means that source $k$ is dominant in the \gls{tf}-bin $(t,f)$.
Giving noise the index $0$ we have:
\begin{align}
	\label{eq:spp_variable}
	\vect{y}_{t,f} = \begin{cases} \vect{x}_{k,t,f}  & z_{k,t,f} = 1; \; k=1,\ldots , K \\  
    \vect{n}_{t,f} & z_{0,t,f} = 1.\end{cases}
\end{align}
This latent variable can be estimated using the \gls{EM} algorithm applied to a spatial mixture model for $\vect{y}_{t,f}$ \cite{ito2016complex}:%\cite{Higuchi_2016}:
\begin{align}
    p(\vect{y}_{t,f}) = \sum_{k=0}^K \pi_k p(\vect{y}_{t,f};\boldsymbol{\theta}_k),
\end{align}
where $\pi_k$ is the a priori probability that $z_{k,t,f} = 1$, which can be chosen time frame or frequency dependent.
Further, $p(\vect{y}_{t,f};\boldsymbol{\theta}_k)$ is a distribution that models the spatial properties of  $\vect{y}_{t,f}$ if it is from source $k$ \cite{ito2016complex}.
The \gls{EM} algorithm  estimates the posterior probability
\begin{align}
    \label{eq:EM:posterior}
    m_{k,t,f} = \ExpOp[z_{k,t,f}|\vect{y}_{t,f}]\; \in [0,1] .
\end{align}
This, or a quantized version of it, is called a mask.
\Cref{fig:masks} shows example masks computed from a segment of audio, which consists of partially overlapping speech of two speakers and additive noise.

With the estimated masks, the \glspl{SCM} can be estimated as follows:
%\difftwo{\begin{align}
%	\label{eq:scm_speech_blockestimate}
%	\scmtrue_{\vect{x}_k,f} &= \sum_{\tau=1}^T \frac{1}{\sum_{\tau'=1}^T m_{k,\tau',f}} m_{k,\tau,f}\vect{y}_{\tau,f} \vect{y}_{\tau,f}^{\HT}, \quad k=1, \ldots , K\\
%	\scmtrue_{\vect{n},f} &= \sum_{\tau=1}^T \frac{1}{\sum_{\tau'=1}^T m_{0,\tau',f}} m_{0,\tau,f}\vect{y}_{\tau,f} \vect{y}_{\tau,f}^{\HT}.
%\end{align}}
\begin{align}
	\label{eq:scm_speech_blockestimate}
	\scmest_{\vect{x}_k,f} &= \sum_{\tau=1}^T \frac{1}{\sum_{\tau'=1}^T \hat{m}_{k,\tau',f}} \hat{m}_{k,\tau,f}\vect{y}_{\tau,f} \vect{y}_{\tau,f}^{\HT}, \quad k=1, \ldots , K\\
	\scmest_{\vect{n},f} &= \sum_{\tau=1}^T \frac{1}{\sum_{\tau'=1}^T \hat{m}_{0,\tau',f}} \hat{m}_{0,\tau,f}\vect{y}_{\tau,f} \vect{y}_{\tau,f}^{\HT}.
\end{align}
With the \gls{SCM} estimates, the beamformer coefficients of the various statistically optimum beamformers can be readily computed.
Applying the beamformer as in Eq.~(\ref{eq:mvdr}) realizes denoising and/or source separation.

Another important class of model-based source separation technique is  \gls{BSS} based on \gls{ICA}.
It estimates spatial filters that separate a given number of sources only from the observation, assuming its underlying physical model and statistical independence between the sources.
Many techniques have been developed for speech source separation based on these principles\cite{gannot2017consolidated,sawada2020apsipa}.

%%%%%%%%%%%%%%%%%%%%%%%%%%%%%%%%%%%%%%%%%%%%%%%%%%
\subsubsection{Dereverberation}
\label{sec:model_dereverberation}
%%%%%%%%%%%%%%%%%%%%%%%%%%%%%%%%%%%%%%%%%%%%

The goal of dereverberation is the deconvolution of the speech signal to remove the effect of the reverberant environment.
We here introduce two model-based methods, the first is based on a transversal and the second on an autoregressive model of the reverberated signal.

\Cref{eq:ctf} can be expressed in vector form (let $K=1$ for ease of exposition):
\begin{align}
    \label{eq:reverb_vectorform}
    \vect{y}_{t,f} &= \breve{\vect{H}}_{t,f}\breve{\vect{s}}_{t,f} + \vect{n}_{t,f}
\end{align}
where 
\begin{align}
    \label{eq:state_equation}
    \breve{\vect{H}}_{t,f} &= \begin{bmatrix} \vect{h}_{0,f} & \vect{h}_{1,f} & \cdots & \vect{h}_{L-1,f} \end{bmatrix} \in \mathbb{C}^{D\times L}\\
    \label{eq:observation_equation}
    \breve{\vect{s}}_{t,f} &= \begin{bmatrix} s_{t,f} & s_{t-1,f} & \cdots & s_{t-L+1,f} \end{bmatrix}^\T \in \mathbb{C}^L.
\end{align}
If we model $s_{t,f}$ as complex Gaussian random variable, $p(s_{t,f};\lambda_{t,f}^{\text{(s)}}) = \mathcal{CN}(s_{t,f}; 0, \lambda_{t,f}^{\text{(s)}})$, we arrive at a Gaussian random process for $\breve{\vect{s}}_{t,f}$ with the following state equation
\begin{align}
    \breve{\vect{s}}_{t,f} &= \vect{A}_{t,f} \breve{\vect{s}}_{t-1,f} + \vect{w}_{t,f}.
\end{align}
The state transition matrix $\vect{A}$ and the system noise vector $\vect{w}$ can readily be identified by comparing this state space model with \cref{eq:observation_equation}.

Based on this view, a dereverberation algorithm has been developed, which consists of an \gls{EM} algorithm with parameter estimation in the M-step, while the E-step is a Kalman filter for the actual enhancement \cite{schwartz2014online}.

The second dereverberation algorithm, the \gls{WPE} method \cite{Yoshioka_2011}, approximates the reverberated speech, \cref{eq:reverb_only}, by an \gls{AR} model (also referred to as a linear prediction model):
\begin{align}
	\label{eq:reverb_only_ar}
	\vect{y}_{t,f} &=  \vect{x}_{t,f}^{\text{(early)}} + \vect{x}_{t,f}^{\text{(late)}} \nonumber \\
	&= \sum_{\tau=0}^{\Delta -1} \vect{h}_{\tau,f}s_{t-\tau,f} +
	\sum_{\tau=\Delta}^{L -1} \vect{G}_{\tau,f}\vect{y}_{t-\tau,f},
\end{align}
where the matrix $\vect{G}_{t,f} \in \mathbb{C}^{D\times D}$ is a matrix with the coefficients of the \gls{AR} process $\vect{y}_{t,f}$.
However, speech is an \gls{AR} process by itself.
To not destroy that property and capture only the unwanted reverberation in $\vect{x}_{t,f}^{\text{(late)}}$, the lag $\Delta$ is introduced.
Thus, the dereverberated speech can be estimated via \cite{Yoshioka_2011}
\begin{align}
	\label{eq:wpe}
	\hat{\vect{x}}_{t,f}^{\text{(early)}} 
	&= \vect{y}_{t,f} - \sum_{\tau=\Delta}^{L' -1} \hat{\vect{G}}_{\tau,f}\vect{y}_{t-\tau,f},
\end{align}
where $\hat{\vect{G}}$ is an estimate of $\vect{G}$.
WPE maximizes the likelihood of the model under the assumption that $\vect{x}_{t,f}^{\text{(early)}}$ is a realization of a zero-mean complex Gaussian process with an unknown channel independent time-varying variance $\lambda_{t,f}$:
\begin{align}
	\label{eq:wpe_compGauss}p(\vect{x}_{t,f}^{\text{(early)}};\lambda_{t,f}) = \mathcal{CN}(\vect{x}_{t,f}^{\text{(early)}}, 0, \lambda_{t,f}\vect{I}).
\end{align}
Here, $\vect{I}$ is the identity matrix.
An iterative estimation algorithm has been developed for alternately estimating $\vect{G}$ and $\lambda_{t,f}$ \cite{Yoshioka_2011}.

%%%%%%%%%%%%%%%%%%%%%%%%%%%%%%%%%%%%%%%%%%%%%%%%%%
\subsubsection{Remarks}
\label{sec:model_remarks}
%%%%%%%%%%%%%%%%%%%%%%%%%%%%%%%%%%%%%%%%%%%%

The presented model-based approaches to beamforming or dereverberation estimate their parameters from the microphone signal to be enhanced.
Since there is no separate training phase, there is no risk of a train-test mismatch, i.e., that the training data have different statistics than the data to be enhanced.
In fact, they can adapt to a new acoustic environment with only a few seconds of observed data.

Since the model parameters to be estimated are not directly observable, 
the \gls{EM} is a good starting point to derive an estimation algorithm.
It is an iterative batch algorithm and thus not readily usable for online, low-latency processing.
However, recursive \gls{EM} formulations exist, which then allow low-latency processing \cite{titterington1984recursive}.

Model-based techniques have been shown to be very robust.
For example, the spatial mixture model-based source activity estimation \cite{ito2016complex} and the \gls{WPE} dereverberation \cite{Yoshioka_2011} have been chosen for the CHiME-6\diffsimple{}{ \cite{Watanabe2020CHiME6}} and CHiME-7\diffsimple{}{ \cite{cornell23_chime7}} challenge baseline systems.
This data set on dinner party transcription is known for its extremely challenging acoustic conditions.
The high robustness of the model-based approach is probably due to the physical and statistical assumptions that are implemented in the model with relatively few parameters.

Given an underlying physical model, the derived algorithms often enjoy explainability: the effect of certain parameters on the outcome can often be well predicted.
For example, for the above beamformers, the signal-to-noise ratio gain through beamforming can be computed, and the effect of the noise statistics on that gain can be assessed analytically.
In case of Bayesian estimators one may even be able to predict the reliability of the estimate.

The above examples can only give a glimpse into the rich literature on model-based enhancement.
They have been chosen to highlight the typical properties of model-based enhancement.
We are aware of the fact that many important techniques had to be left out from our discussion due to space limitations.

%%%%%%%%%%%%%%%%%%%%%%%%%%%%%%%%%%%%%%%%%%%%%%%%%%
\subsection{Data-driven parameter estimation and enhancement}
\label{sec:data_driven}
%%%%%%%%%%%%%%%%%%%%%%%%%%%%%%%%%%%%%%%%%%%%

With data-driven techniques, we refer to machine-learning-based methods.
They take a different route than the model-based approach: instead of defining a physical model from which the enhancement operation can be derived, the enhancement function is estimated from examples of input-output pairs $(\vect{y}, s)$, with  $\vect{y}$ being the input to the enhancement stage and $s$  being the desired output.

The enhancement operator is nowadays realized by a \gls{DNN}.
Clearly, a \gls{DNN} is also a model, although not physically inspired.
%Its architecture and size determine the space of mapping functions that can be realized, i.e., the hypothesis space.
The \gls{DNN} has in general many more parameters than the physically and statistically inspired models discussed above.
To discern it from those, we won't call it a ``model'' in the following.

%%%%%%%%%%%%%%%%%%%%%%%%%%%%%%%%%%%%%%%%%%%%%%%%%%
\subsubsection{Denoising and source separation}
\label{sec:data_denoising}
%%%%%%%%%%%%%%%%%%%%%%%%%%%%%%%%%%%%%%%%%%%%

There are many examples of data-driven approaches to speech enhancement \cite{wang2018supervised}.
The vast majority of neural networks are real-valued machines.
However, the \gls{STFT} representations are complex-valued variables.
Thus, neural networks were initially not good at representing phase information.
Consequently, the first neural regression-based enhancement techniques only estimated the magnitude spectrum of the desired signal from the \gls{STFT} magnitude representation of the input signal, disregarding the phase \cite{xu2014regression}.
Later it was found that stacking the real and imaginary components of the input \gls{STFT} was a simple but effective means to represent the complex-valued input signal.
So-called complex spectral mapping computes the target \gls{RI} components based on the \gls{RI} components of the input signal with a regression neural network \cite{wang2020complex}.
Or the network is trained to estimate a complex-valued mask to do so-called complex ratio masking \cite{liu2019divide}.
%There is no principled advantage of one over the other.

An alternative way to avoid the issue with representing complex-valued variables is to do ``time-domain'' processing.
Here, the \gls{STFT} and inverse \gls{STFT} are replaced by learnt encoder-decoder modules  \cite{luo2019conv}.
Using very short analysis windows of, e.g., \SI{2}{ms}, these techniques achieved excellent single-channel source separation performance.

%\Gls{DNN}-based approaches were celebrated for achieving unprecedented single-channel speech separation performance
%\cite{kolbaek2017multitalker, hershey2016deep}....

\Glspl{DNN} were \diff{}{initially} also poor at modeling spatial information available in multi-channel data, because this manifests itself in phase differences between the frequency domain representations of the channels of a compact microphone array, see \cref{eq:anechoic}.
This was in part overcome by computing spatial features, such as inter-channel phase differences between the \glspl{STFT} of the microphone signals \cite{wang2018multi} and appending these features with spectral features at the input of the \gls{DNN}.
\diff{}{Furthermore,} the above technique of stacking real and imaginary parts to a real-valued vector of double the size proved to be effective to capture spatial cues as well\diffsimple{}{ \cite{wang2021multi}}.

For processing in the \gls{STFT} domain, a typical representation of multi-channel data at the input to the network is a three-dimensional tensor with time, frequency and channel axes, where the latter consists of $2D$ components, capturing the real and imaginary values of the signals in the $D$ microphone channels.
This is complemented by a fourth dimension, the batch dimension.

An example network architecture for multi-channel speech enhancement consists of alternating layers for temporal processing and spectral processing \cite{wang2022tfgridnet_2022}.
Temporal processing is achieved by a recurrent network layer (e.g., a \gls{BLSTM} layer) whose sequence axis is the time axis and which operates independently for each frequency bin.
Spectral processing is done with a\diff{ recurrent (BLSTM) network layer}{ network layer (e.g., a recurrent BLSTM layer)} that operates along the frequency axis, for each time frame separately.
This is sometimes called ``fullband processing'' and is known to be particularly effective in capturing spatial information, because the \gls{DOA} results in a characteristic phase change pattern along the frequency axis.

For denoising, the network predicts masks for the speech and noise.
Since the two have distinct spectral patterns, a \gls{DNN} can learn easily which of its outputs corresponds to the speech and which to the noise masks.
However, for source separation, we should output masks for each speaker.
Speech signals have similar spectral patterns, which causes a permutation ambiguity, i.e., the mapping between \gls{DNN} outputs and sources are arbitrary.
\Gls{PIT} was introduced as a way to circumvent this ambiguity to allow training \glspl{DNN} for speech separation \cite{kolbaek2017multitalker}.
The idea is simply to compute the training loss with the optimal permutation of the sources with respect to the loss.
\Gls{PIT} is used in most \gls{DNN}-based speech separation approaches.

\begin{figure}
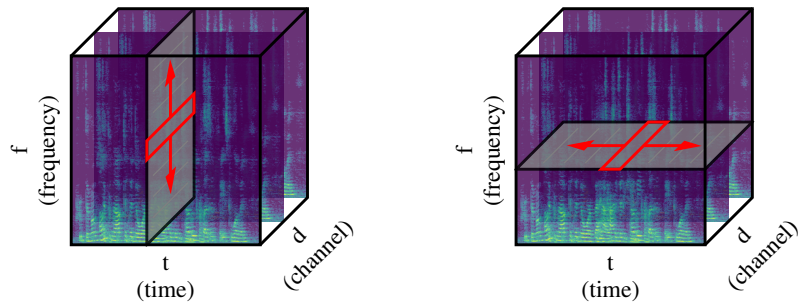

    \begin{center}
    \parbox{0.35\textwidth}{
        \newcommand{\sliceopacity}{0}
        \input{figures/fullband_subband.tex}
    }
    \parbox{0.35\textwidth}{
        \newcommand{\sliceTwoopacity}{0}
        \input{figures/fullband_subband.tex}
    }
        \caption{An illustration of spectral (fullband) processing within a frame (left) and temporal (subband) processing across frames (right).
        The red arrows indicate the sequence axis.}
        \label{fig:fullband_subband}
    \end{center}
\end{figure}

%%%%%%%%%%%%%%%%%%%%%%%%%%%%%%%%%%%%%%%%%%%%%%%%%%
\subsubsection{Dereverberation}
\label{sec:data_dereverberation}
%%%%%%%%%%%%%%%%%%%%%%%%%%%%%%%%%%%%%%%%%%%%

As late reverberation can be viewed as an additive distortion of the desired signal, see \cref{eq:ctf_2}, \gls{DNN}-based networks for dereverberation are similar to those for denoising.
One of the earliest works was \cite{weninger2014deep}, where a recurrent neural network was employed in an autoencoder structure with the magnitude spectrum of the reverberated speech at the input and the magnitude spectrum of the non-reverberant speech as the training target at the output.
Many more deep learning-based systems have been proposed, some of which are capable of jointly carrying out dereverberation, denoising, and source separation 
\cite{wang2022tfgridnet_2022}.
%\cite{Shengkui2023icassp,wang2022tfgridnet_2022}.%

%%%%%%%%%%%%%%%%%%%%%%%%%%%%%%%%%%%%%%%%%%%%%%%%%%
\subsubsection{Remarks}
\label{sec:data_remarks}
%%%%%%%%%%%%%%%%%%%%%%%%%%%%%%%%%%%%%%%%%%%%%%%%%%

Parameter estimation of \glspl{DNN} is also guided by a statistically motivated objective function, typically the cross entropy in case of a network for classification (e.g., for mask estimation), which is an instance of \diff{Maximum Likelihood}{maximum likelihood} estimation.
In this sense, there is some similarity to the model-based approaches that are derived to be the solution to a statistical optimization problem.
However, model-based beamforming can choose among several objective functions, of which the constrained optimization leading to the \gls{MVDR} criterion is a particularly interesting one, because, by definition, it does not introduce speech distortions.
No such distortion-less criterion is known for \glspl{DNN}.

However, this is a rather theoretical argument, because experimentation tells that \glspl{DNN} are excellent enhancement machines.
No simplifying modeling assumptions, or at least no obvious simplifying assumptions are made in the data-driven approach.
Consequently, very good performance has been reported for many enhancement tasks, and purely data-driven techniques are often at the top of leaderboards for single-channel processing.%
% Consequently,  very good performance has been reported for many enhancement tasks, and purely data-driven techniques are often at the top of leaderboards.
\footnote{See, e.g., \url{https://paperswithcode.com/sota/speech-enhancement-on-deep-noise-suppression},
\url{https://paperswithcode.com/sota/speech-separation-on-wsj0-2mix} or
\url{https://paperswithcode.com/sota/speech-separation-on-libri2mix}}\label{footnote_leaderboards}
Similarly, data-driven approaches are also becoming more competitive for multi-channel processing, as discussed in \cref{sec:beamforming_case_study}, although it is more difficult to measure because results are often reported on different corpora.

%Due to the many parameters of the \gls{DNN}, the amount of training data considered necessary to obtain good enhancement performance is significant.
% \inred{Can we give some concrete numbers?}.
% To avoid costly data recording and annotation campaigns, researchers regularly resort to simulation and data augmentation techniques.
% For example, \glspl{RIR} are generated by simulation tools using the image source method and variants of \cite{habets2006room}.
% For the supervised training of enhancement networks, so-called paired data, which consist of pairs of clean and degraded versions of the same utterance, are required.
% The pairs are generated by artificially introducing distortions to the clean signal because real recordings of paired data are hard to obtain: one would require a special recording setup to capture both the clean source speech signal and the microphone signal at the same time.
% Data augmentation refers to expanding the training set by artificially modifying/distorting the input data and letting the network learn to disregard those modifications.

Due to the many parameters of the \gls{DNN}, the amount of training data considered necessary to obtain good enhancement performance is significant.
%\inred{Can we give some concrete numbers?}.
For the supervised training of enhancement networks, so-called paired data, which consist of pairs of clean and degraded versions of the same utterance, are required.
Collecting such paired data for real recordings involves capturing both the clean source speech signal and the microphone signals simultaneously, which requires a special recording setup and is often difficult, if not impossible.
To avoid costly data recording and annotation campaigns, researchers regularly resort to simulation and data augmentation techniques.
Simulated microphone signals can be generated following \cref{eq:time_domain_convolution} using a dataset of clean speech signals and noise recordings.
Besides, \glspl{AIR} can also be artificially generated by simulation tools using the image source method and variants of it.
In addition, data augmentation is also frequently used to expand the training set by artificially modifying/distorting the input data and letting the network learn to disregard those modifications.
With these techniques, arbitrarily large amounts of training data can be generated.

However, these artificial degradations of clean speech can only approximate the true degradation signals experience in a real environment\diffsimple{}{ \cite{zmolikova2023masked,wisdom2020mom}}. One known issue with artificial data is that they almost exclusively use time-invariant \glspl{AIR}, while in a real recording, the \gls{AIR} is time-varying due to movements of the speaker, even if those are only small head movements.
\difftwo{To mitigate these approximations, several works have attempted to exploit real recordings for training or adapting the networks using, e.g., unsupervised learning approaches\mbox{ \cite{Drude_2019b,wisdom2020mom,zmolikova2023masked}}.}{In an attempt to be able to train on real data, for which the separated source signals are unavailable, mixture invariant training (MixIT) has been proposed, which is completely unsupervised \cite{wisdom2020mom}: Here, existing mixtures are mixed together. The algorithm then attempts to separate the mixture of mixtures into the individual source signals such that, if remixed, the original mixture of mixtures can be retained. Other unsupervised domain adaptation techniques have been developed, that require only the degraded signal from the target environment and not paired degraded - clean recordings \cite{zmolikova2023masked}.}

%Data-driven techniques are only as good as representative the training data is of the test data.
\difftwo{Data-driven techniques}{While these attempts diminish the train-test mismatch, they cannot fully close the gap, and it remains true that data-driven techniques} are only good when the training and test data characteristics are relatively similar.
If the statistics of test data do not match that of the training data, then performance can be poor.
Thus, the quality of data-driven enhancement crucially depends on the availability of appropriate training data.
%There is also considerable work on domain adaptation techniques.
% Here, the goal is to adapt/transfer a well-trained enhancement network to new domain with only a few in-domain training data \inred{more on this?}.

But even if a large amount of representative training data is used, there is still the risk that some test data elicit the network in regions of feature space where it is not well trained.
It is a known issue that small modifications of the input data can lead to surprisingly poor output\difftwo{}{ \cite{VINCENT2017535}}.
Thus, data-driven techniques are generally viewed to be ``black boxes'', with sometimes surprising and unexplainable results.

Finally, \gls{DNN}-based enhancement tends to be more compute-intensive and memory-demanding than model-based enhancement, but novel frameworks and specialized hardware diminish this drawback.

%%%%%%%%%%%%%%%%%%%%%%%%%%%%%%%%%%%%%%%%%%%%%%%%%%
%\subsection{Hybrid speech enhancement}
%\label{sec:hybrid}
%%%%%%%%%%%%%%%%%%%%%%%%%%%%%%%%%%%%%%%%%%%%%%%%%%

%%%%%%%%%%%%%%%%%%%%%%%%%%%%%%%%%%%%%%%%%%%%%%%%%%
\section{Hybrid speech enhancement}
\label{sec:taxonomy}
%%%%%%%%%%%%%%%%%%%%%%%%%%%%%%%%%%%%%%%%%%%%

If a physical and/or statistical model exists that reflects reality well, whose parameters are easy to obtain, and which leads to effective algorithms, there is no need to switch to data-driven methods.
On the other hand, if the nature of the degradation and the statistics of the involved signals are so complex that they cannot be well represented by a computationally accessible model, and if appropriate training data are available or can be simulated, then data-driven methods are the preferred choice.

Often, however, the truth lies somewhere in-between.
From the preceding discussion, it becomes obvious that model-based and data-driven methods have different and often complementary properties.
It is thus natural to search for combinations of the two to have the best of both worlds.
Such hybrid systems have been proposed for many enhancement tasks.
% We here offer a taxonomy and give representative examples.

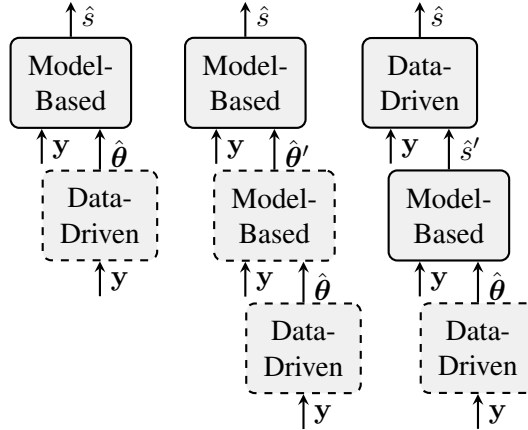
\begin{figure}
    \centering
    \begin{tikzpicture}
    \linespread{1.}

    \newcommand{\shift}{1em}  % distance of an arrow to the center of a block
    \newcommand{\gap}{2em}  % gap between two processing blocks
    \newcommand{\figuregap}{6em}
    \newcommand{\ModelBased}{Model-\\Based}
    \newcommand{\DataDriven}{Data-\\Driven}

    \begin{scope}[xshift=-\figuregap]
        \node[block, enhBlock, align=center] (AenhModel) at (0, 0) {\ModelBased};
        \node[block, estBlock, align=center, anchor=north, dnn] (AestDNN) at ($(AenhModel.south)+(\shift, -1.4em)$) {\DataDriven};
        
        \draw[arrow] (AestDNN) -- (AenhModel.south-|AestDNN) node[below right] (theta) {$\hat{\vect{\theta}}$};
        \draw[arrow] ($(AenhModel.south)+(-\shift, 0)$) +(0, -1.2em) -- +(0, 0) node[below right] {$\vect{y}$};
        
        \draw[arrow] (AenhModel.north) node[above right] {$\hat{s}$} -- +(0,1.2em);
        \draw[arrow] (AestDNN.south) node[below right] {$\vect{y}$} +(0,-1.2em) -- +(0,0);
    \end{scope}
    
    \begin{scope}[xshift=0]
        \node[block, enhBlock, align=center] (BenhModel) at (0, 0) {\ModelBased};
        \node[block, estBlock, align=center, anchor=north] (BenhModelTwo) at ($(BenhModel.south)+(\shift, -1.4em)$) {\ModelBased};
        \node[block, estBlock, align=center, anchor=north, dnn] (BestDNN) at ($(BenhModelTwo.south)+(\shift, -1.4em)$) {\DataDriven};
        
        \draw[arrow] (BenhModelTwo) -- (BenhModel.south-|BenhModelTwo) node[below right] (theta) {$\hat{\vect{\theta}}\smash{'}$};
        \draw[arrow] (BestDNN) -- (BenhModelTwo.south-|BestDNN) node[below right] (theta) {$\hat{\vect{\theta}}$};
        \draw[arrow] ($(BenhModel.south)+(-\shift, 0)$) +(0, -1.2em) -- +(0, 0) node[below right] {$\vect{y}$};
        \draw[arrow] ($(BenhModelTwo.south)+(-\shift, 0)$) +(0, -1.2em) -- +(0, 0) node[below right] {$\vect{y}$};
        
        \draw[arrow] (BenhModel.north) node[above right] {$\hat{s}$} -- +(0,1.2em);
        \draw[arrow] (BestDNN.south) node[below right] {$\vect{y}$} +(0,-1.2em) -- +(0,0);
    \end{scope}
    
    \begin{scope}[xshift=\figuregap]
        \node[block, enhBlock, align=center, dnn] (CenhModel) at (0, 0) {\DataDriven};
        \node[block, enhBlock, align=center, anchor=north] (CenhModelTwo) at ($(CenhModel.south)+(\shift, -1.4em)$) {\ModelBased};
        \node[block, estBlock, align=center, anchor=north, dnn] (CestDNN) at ($(CenhModelTwo.south)+(\shift, -1.4em)$) {\DataDriven};
        
        \draw[arrow] (CenhModelTwo) -- (CenhModel.south-|CenhModelTwo) node[below right] (theta) {\diffsimple{$\hat{\vect{\theta}}\smash{'}$}{$\hat{s}\smash{'}$}};
        \draw[arrow] (CestDNN) -- (CenhModelTwo.south-|CestDNN) node[below right] (theta) {$\hat{\vect{\theta}}$};
        \draw[arrow] ($(CenhModel.south)+(-\shift, 0)$) +(0, -1.2em) -- +(0, 0) node[below right] {$\vect{y}$};
        \draw[arrow] ($(CenhModelTwo.south)+(-\shift, 0)$) +(0, -1.2em) -- +(0, 0) node[below right] {$\vect{y}$};
        
        \draw[arrow] (CenhModel.north) node[above right] {$\hat{s}$} -- +(0,1.2em);
        \draw[arrow] (CestDNN.south) node[below right] {$\vect{y}$} +(0,-1.2em) -- +(0,0);
    \end{scope}
\end{tikzpicture}
    \caption{%
Three classes of hybrid approaches (from left to right): data-driven parameter estimation for model-based enhancement; combined model-based and data-driven parameter estimation for model-based enhancement; joint model-based and data-driven approach for the enhancement operation.}
    \label{fig:taxonomy_hybrid}
\end{figure}

%The linearity assumption and the additivity of the noise underlying the signal models of \cref{sec:signal_models} are well justified.
% Thus, many enhancement solutions have been developed based on those models.
% However, those enhancement systems will, in almost all cases, have parameters whose values are not known in advance.
% These can be parameters of the involved random processes, the underlying physical model, or the parameters of the enhancement system itself.

%As we showed in \cref{fig:enhancement_system}, the enhancement process can be divided into two operations: parameter estimation and enhancement.
In this section, we describe three classes of hybrid approaches that differ in how they use model-based and data-driven methods for parameter estimation or for enhancement, see \cref{fig:taxonomy_hybrid}.
The first exploits a data-driven approach to estimate the parameters of a model-based enhancement method.
The second combines data-driven and model-based approaches for estimating the model parameters.
Finally, the last combines model-based and data-driven approaches for the enhancement operation itself.
We will also discuss how to train a hybrid system and study the gradual replacement of model-based components by learnable modules with the example of acoustic beamforming.
%In \cref{fig:enhancement_system}, we have seen two variants for estimating the enhancement system's parameters.
% We have discussed an example of the first (estimation of the parameters from the signal to be enhanced) already in \cref{sec:modelbased}.
% Here, we will discuss an example of the second option (estimation of the parameters based on training data).
% Because of the complementary properties of the two variants, we will then describe combinations of the two.
% There are also combinations of model-based and data-driven approaches for the enhancement operation itself, of which we will present an example as well.
% We will also discuss how to train hybrid system and study the gradual replacement of model-based components by learnable modules with the example of acoustic beamforming.
%%%%%%%%%%%%%%%%%%%%%%%%%%%%%%%%%%%%%%%%%%%%%%%%%%
%\subsection{Data-driven parameter estimation for model-based enhancement}
\subsection{Data-driven estimation of model parameters}
\label{sec:joint_model_data_driven_enhancement}
%%%%%%%%%%%%%%%%%%%%%%%%%%%%%%%%%%%%%%%%%%%%

In \cref{sec:modelbased}, the source activity masks were estimated via the posterior class probabilities of a spatial mixture model\diffsimple{}{, as shown in \cref{eq:EM:posterior}}. With those, the \glspl{SCM} were estimated, from which eventually the beamformer coefficients were obtained.
Alternatively, the masks can be estimated by a \gls{DNN} in a supervised learning approach, where the noisy speech is at the input and the class affiliation $\vect{z}_{t,f}$ or quantities derived from it are the training targets \cite{Heymann_2016, erdogan16_interspeech}.
In its simplest variant, the training objective is a binary cross entropy criterion, which computes, for each frequency $f$, time frame $t$ and source $k$, the cross entropy between the network output $\hat{z}_{k,t,f}$ and the ground truth $z_{k,t,f}$.
%The network architecture often contains a recurrent or convolutive network layer to capture temporal information.
Because all \gls{STFT} bins $f=0,1,\ldots, F-1$ are used as inputs, the network can effectively learn the spectro-temporal patterns of speech and noise.
On the contrary, the spatial mixture model based estimator \cite{ito2016complex} %\cite{Higuchi_2016} 
operates on each frequency bin independently, which introduces a frequency permutation problem that needs to be addressed.
%If inter-channel phase differences are used as additional cross-channel features at the input, the network can also incorporate spatial cues for mask estimation \cite{wang2018multi}.

Many variants of this basic approach have been proposed.
A particularly interesting one is concerned with speech enhancement in dynamic acoustic environments, where the signal statistics change over time.

In \cref{eq:scm_speech_blockestimate}, a time-invariant \gls{SCM} is estimated over a block of $T$ frames.
However, if speakers move, the \glspl{SCM} are time-varying, and we need to introduce a tracking mechanism.
In its simplest form, this is achieved by a first-order recursive filter %\cite{togami2019simultaneous}
\begin{align}
	\label{eq:scm_speech_recursiveestimate}
	\scmest_{\vect{x}_k,t,f} &= \beta \scmest_{\vect{x}_k,t-1,f} + \instscmest_{\vect{x}_k,t,f} 
	= \sum_{\tau = 1}^t \beta^{t-\tau}\instscmest_{\vect{x}_k,\tau,f},
\end{align}
where 
\begin{align}
	\instscmest_{\vect{x}_k,t,f} = m_{k,t,f} \vect{y}_{t,f} \vect{y}_{t,f}^{\HT}
\end{align}
is the \gls{ISCM}, and $0 < \beta < 1$ is a forgetting factor, which gives exponentially less weight to older \glspl{ISCM}.
However, setting the forgetting factor is challenging because its optimal value depends on the rate of change of the acoustic environment.

We can express the different \gls{SCM} computations by writing \cite{Ochiai_2023}
\begin{align}
	\label{eq:scm_speech_attention}
	\scmest_{\vect{x}_k,t,f} &= \sum_{t'=1}^T c_{\vect{x}_k,t,t'} \instscmest_{\vect{x}_k,t',f},
\end{align}
where the definition of $c_{\vect{x}_k,t,t'}$ can be inferred by comparing \cref{eq:scm_speech_attention} with \cref{eq:scm_speech_blockestimate} and \cref{eq:scm_speech_recursiveestimate}.
These coefficients control the range for accumulating the statistics for computing the \gls{SCM}.
The same generalization can also be applied to the estimation of the \gls{SCM} of the noise.
\cref{eq:scm_speech_attention} has a similar form as the well-known (self) attention mechanism of deep learning, and attention networks can thus be used to find optimal weights for the \gls{SCM} estimation in dynamic acoustic environments \cite{Ochiai_2023}.

%%%%%%%%%%%%%%%%%%%%%%%%%%%%%%%%%%%%%%%%%%%%%%%%%%
%\subsection{Joint model-based and data-driven parameter estimation for model-based enhancement}
\subsection{Combined data-driven and model-based estimation of model parameters}
\label{sec:joint_model_data_driven_parameter}
%%%%%%%%%%%%%%%%%%%%%%%%%%%%%%%%%%%%%%%%%%%%

The two methods for \gls{tf} mask estimation, spatial mixture models and the \gls{DNN}-based approach, exploit different signal properties: The spatial clustering approach utilizes the spatial diversity of the sources, while the \gls{DNN} is trained to learn the typical spectro-temporal patterns of speech and noise.
The former does unsupervised parameter learning on the utterance to be enhanced, while the latter employs a training stage.

It is thus natural to combine the two strategies.
%In one approach, the trained DNN-based mask estimator is applied to the test utterance to obtain the initial values of the masks.
% Those are then refined with the EM algorithm as explained in \cref{sec:modelbased} \cite{nakatani2017integrating}.
One approach integrates the two mask estimators by using the masks obtained by a \gls{DNN} as a priori probability of the spatial mixture model.
The \gls{EM} algorithm refines the masks as the posterior probability of the model \cite{nakatani2017integrating}.

\difftwo{}{While this technique carries out unsupervised adaptation of the DNN-computed masks on the test utterance, the unsupervised learning capability of the spatial mixture model can also be used in the training stage: 
%the separated signals 
the posterior class probabilities
computed by the mixture model can be taken as training targets for the supervised training of the DNN-based mask estimator. This has led to better separation performance, because the DNN-based mask estimator treats all frequencies jointly, unlike the spatial mixture model \cite{Drude_2019b}. Furthermore, this system can perform source separation on single-channel inputs, despite having learned how to do so using multi-channel recordings \cite{Tzinis_2019}.}

%In another approach, the representation of spectro-temporal properties computed by a DNN, and the spatial properties present in the vector $\vect{y}_{t,f}$ are considered two observations of the same underlying latent variable $z_{k,t,f}$.
%Its value can be estimated by an EM algorithm that integrates both observation models \cite{drude2019integration}.
%This approach enables the unsupervised learning of the DNN part with the help of the model-based approach.

The \gls{tf} masks can also be estimated in a two-step procedure, starting with a source activity variable $m_{k,t}$ with time resolution, that is refined to \gls{tf} resolution $m_{k,t,f}$ in the second stage.
If the first stage, the diarization stage, which estimates the activity of the individual speakers, is realized with a \gls{DNN}, and the second with a model-based approach, this is another example of joint model-based and data-driven speech activity estimation.
In its simplest variant, the \gls{DNN} in the first stage is trained to detect speech overlap, where concurrent speakers are active, and exclude those segments in the second stage.
The remaining segments in the second stage can then be used to compute the beamformer coefficients for extracting each of the relevant source signals \cite{chazan2018lcmv}.
But there are also neural diarization methods that can cope with speech overlap and identify each of the simultaneously active speakers.
A particularly powerful one, the \gls{TSVAD}, can identify the speakers active in a segment of overlapped speech \cite{medennikov2020target}.
The refinement of the computed diarization information $m_{k,t}$ to \gls{tf} resolution can be achieved with the very same spatial mixture model mentioned earlier in \cref{sec:model_beamforming}.
Here, the diarization information is utilized to constrain (``guide'') the estimation of the posterior: $m_{k,t,f}$ is allowed to have non-zero values only in time frames, where $m_{k,t}$ is \diff{non-negative}{not zero} \cite{boeddeker2018chime5}, %
\diffsimple{}{%
i.e., when the target speaker is found to be active by the diarization front-end. %
This modification can be simply implemented by replacing \cref{eq:EM:posterior} with
\begin{align} \label{eq:gss}
    m_{k,t,f}&=\frac{m_{k,t}\ExpOp[z_{k,t,f}|\vect{y}_{t,f}]}{\sum_{\tilde{k} = 1}^{K} m_{\tilde{k},t}\ExpOp[z_{\tilde{k},t,f}|\vect{y}_{t,f}]}
\end{align}
in the EM algorithm of the model-based approach.
Thus, the mask estimation shown in \cref{eq:gss} combines model-based and data-driven methods.
% Here, $\ExpOp[z_{k,t,f}|\vect{y}_{t,f}]$ can be estimated based on the EM algorithm, with slight modifications from the model-based approach.
}
This approach, called \gls{GSS}, has turned out to be very robust and was used in most systems, including the top ones, submitted to the CHiME-6\diffsimple{}{ \cite{Watanabe2020CHiME6}} and CHiME-7\diffsimple{}{ \cite{cornell23_chime7}} challenges on dinner party transcription.

There are many more examples of joint \gls{DNN} and model-based parameter estimation.
In the context of \gls{WPE}, for example, the iterative estimation of the model parameters $\lambda_{t,f}$ (power spectral density of clean speech) and of the \gls{AR} parameter matrix $\hat{\vect{G}}_{t,f}$\diffsimple{}{, defined in \cref{eq:reverb_only_ar,eq:wpe_compGauss},}
can be avoided by employing a neural network to estimate $\lambda_{t,f}$ \cite{kinoshita17_interspeech}.
%This not only led to improved performance but also paved the way for an online implementation \cite{Heymann_2018}.

A neural network for target/source \gls{PSD} estimation has also been incorporated into \gls{BSS} algorithms based on \gls{ICA} \cite{Nugraha2016}.
%\cite{Nugraha2016,Makishima2019independent}.
This idea was further pursued for joint dereverberation and source separation using independent vector analysis and vector extraction with a neural source model \cite{Nakatani_2021b,saijo_2022}.

\subsection{Joint model-based and data-driven enhancement}
\label{sec:modeldeficiencies}
%%%%%%%%%%%%%%%%%%%%%%%%%%%%%%%%%%%%%%%%%%%%%%%%%%

Not only have model-based and data-driven parameter estimation complementary advantages and disadvantages, so have the model-based and data-driven enhancement operations.
While the first enjoys statistical optimality under certain assumptions, the second is not constrained by simplifying modeling assumptions.
The preceding discussion has also shown that beamformers are good at exploiting spatial information, while \glspl{DNN} are good at capturing spectro-temporal information.

This  complementarity has been utilized in advanced speech enhancement systems.
A particularly powerful system that prototypically exploits those complementary strengths is the TF-GridNet architecture \cite{wang2022tfgridnet_2022} for joint denoising, dereverberation, and source separation.
Here, the network architecture consists of layers that aim at capturing correlations along the time axis and layers that capture correlations along the frequency axis, similarly as we discussed in \cref{sec:data_denoising} and illustrated in \cref{fig:fullband_subband}.
Further, the self-attention mechanism is introduced to leverage global information across frames.
\difftwo{However, the aggregation of spatial information is achieved with a particular beamformer}{Spatial information is aggregated both by DNNs with multi-channel input and by a beamformer},  a multi-frame Wiener Filter, that is sandwiched between two \glspl{DNN} (\gls{DNN}-BF-\gls{DNN} structure).

%\inred{do you know other examples? ICA/IVA/IVE with DNN-components: mention it here!}

%\inred{If we have one more example, the following can be removed:}
%There are also examples of those combinations for enhancement tasks not covered here.
% Consider \gls{AEC}, to name just one:  The adaptive echo cancellation filter, while being an elegant solution, nevertheless is limited w.r.t. its effectiveness: First, being a linear filter, it is unable to remove nonlinear distortions.
% Those can occur due to loudspeaker imperfections, signal clipping, resonances etc.
% Second, the loudspeaker-to-microphone impulse response may be so long that an adaptation of an echo cancellation filter of the same length would be too slow in tracking the time-varying echo characteristics.
% Then a NN operating on the microphone signal, after the \gls{AEC} has already removed most of the echoes, is an effective means to improve the echo suppression \cite{Haeb-Umbach_2019}.

%%%%%%%%%%%%%%%%%%%%%%%%%%%%%%%%%%%%%%%%%%%%%%%%%%
\subsection{End-to-end optimizaton}
\label{sec:end-to-end}
%%%%%%%%%%%%%%%%%%%%%%%%%%%%%%%%%%%%%%%%%%%%%%%%%%%

A clear deficit of hybrid modeling techniques is that the neural network may be optimized towards a different criterion than the model-based part of the system.
%For example, if the mask estimator is trained with a binary cross entropy criterion to approximate the ideal binary masks of speech and noise, the resulting estimator network may not be optimal for the task of estimating the target speech spectrogram or for the task of speech recognition.
This issue has been addressed in the literature with end-to-end training.
This term means that the whole processing chain is optimized towards a single criterion.
It does not mean that the system has to be a monolithic neural network.

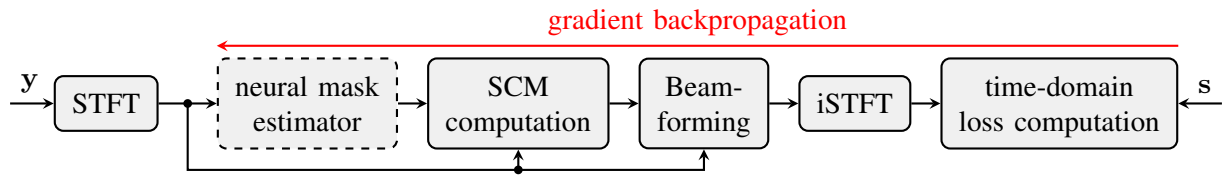
\begin{figure}
	\begin{center}
        {
\linespread{1.}

\begin{tikzpicture}

\node[block] (stft) at (0,0) {STFT};
\node[block,align=center,anchor=west, dnn,estBlock] (nn) at ($(stft.east)+(2em,0)$) {neural mask\\estimator};
\node[block,align=center,anchor=west] (scm) at ($(nn.east)+(1em,0)$) {SCM\\computation};
\node[block,enhBlock,align=center,anchor=west] (bf) at ($(scm.east)+(1em,0)$) {Beam-\\forming};
\node[block,align=center,anchor=west] (istft) at ($(bf.east)+(1em,0)$) {iSTFT};
\node[block,align=center,anchor=west] (loss) at ($(istft.east)+(1em,0)$) {time-domain\\loss computation};

\draw[arrow] (stft.west) + (-1.5em,0) node[above right] {\vect{y}} -- +(0,0);
\draw[arrow] (stft) -- (nn);
\draw[arrow] (nn) -- (scm);
\draw[arrow] (scm) -- (bf);
\draw[arrow] (bf) -- (istft);
\draw[arrow] (istft) -- (loss);
\draw[arrow] ($(stft.east)!1/2!(nn.west)$) node[branch] {} |- ($(nn.south)+(0,-0.7em)$) -| (scm);
\draw[arrow] ($(nn.south-|scm)+(0,-0.7em)$) node[branch] {} -| (bf);
\draw[arrow] (loss.east) + (1.5em,0) node[above left] {\vect{s}} -- +(0,0);

\draw[arrow,red] ($(loss.east|-nn.north)+(0,0.4em)$) -- node[above] {gradient backpropagation} ($(nn.north west)+(0,0.4em)$);

\end{tikzpicture}

}       
	\caption{An example of end-to-end training: Gradient backpropagation from a loss computed in the time-domain through the beamformer to the \gls{DNN} for mask estimation. \cite{boeddeker2021convolutive}.
 }
	\label{fig:end-to-end_example}
	\end{center}
\end{figure}

The end-to-end training criterion can be a signal level metric of the enhanced signal, such as the \gls{SNR} at the output of the beamforming operation.
\Cref{fig:end-to-end_example} provides an example, which we use to illustrate end-to-end optimization \cite{boeddeker2021convolutive}.
Here, the training loss is computed in the time-domain.
Then the gradient is back-propagated through the inverse \gls{STFT} and through the beamforming operation and \gls{SCM} computation all the way to the \gls{DNN} for mask estimation, to learn its parameters.
This requires that \gls{SCM} computation and beamforming operation are differentiable, which they are indeed.

%It displays a neural network-supported beamformer, followed by an automatic speech recognizer \cite{Heymann_2017}.
% The system has two neural networks, one for mask estimation and one for estimating the acoustic model, i.e., the posterior probabilities of speech subword units to be employed for recognizing the spoken words.
% So-called hybrid ASR systems (sorry for the double use of the term ``hybrid'') use a neural network to estimate acoustic subword unit probabilities from the incoming speech.
% That network is trained by minimizing the cross entropy between the estimated subword unit posterior probabilities and the ground truth label sequence.
% In the structure of \cref{fig:beamnet} that loss is back-propagated into the acoustic model network for updating its parameters, but then it is further propagated through the beamforming and \gls{SCM} computation to be finally propagated into the \gls{NN} for mask estimation.
%Also note that the SCM computation and beamforming operations are ``frozen'' (no update of parameters whatsoever).

The end-to-end approach has the advantage that the mask estimator is optimized towards a criterion that is closer to the final objective, such as improving the \gls{SNR}, rather than towards an intermediate criterion.

End-to-end training has also been used to optimize speech enhancement with respect to a loss defined at the output of a downstream system, such as a speech recognizer.
This particular setting has another significant advantage over a separate training of the network in the enhancement stage: the mask estimator training no longer requires paired data.
Instead, the enhancement is trained jointly with the speech recognizer using ``normal'' \gls{ASR} training data, consisting of speech recordings accompanied with their transcription.

Joint optimization of enhancement and \gls{ASR} has been pursued by many researchers, e.g., %\cite{ochiai2017, chang2019mimo}.
\cite{chang2019mimo}.
One should, however, mention that end-to-end training may require a carefully designed training curriculum.
Starting end-to-end training from scratch may result in poor performance, while fine-tuning pre-trained network components with end-to-end training can lead to faster convergence and a better optimum.
%\cite{Heymann_2017}.
Another downside of end-to-end training is that one can lose the modularity of the system.
An enhancement system that is optimized end-to-end with a particular \gls{ASR} model may not perform well when combined with a different \gls{ASR} engine.
%Further, end-to-end training rests upon the assumption that the system components through which the error has to be backpropagated are differentiable.

%%%%%%%%%%%%%%%%%%%%%%%%%%%%%%%%%%%%%%%%%%%%%%%%%%
\subsection{Towards purely data-driven multi-channel enhancement}
\label{sec:beamforming_case_study}
%%%%%%%%%%%%%%%%%%%%%%%%%%%%%%%%%%%%%%%%%%%%%%%%%%%

In the literature, there is a clear trend to replace model-based enhancement with hybrid and even purely data-driven approaches, often termed ``all-neural''.
In fact, even very established representations and algorithms are being replaced by learnable components.

We will show this trend with the example of acoustic beamforming.
The starting point is the configuration of \cref{fig:end-to-end_example}, which consists of a neural network-based mask estimator, while the remaining components to arrive at the enhanced signal are model-based.

% For the computation of the beamformer coefficients, the inverse of $\scm_{\vect{n},f}$ and the \gls{ATF} $\vect{a}_f$ are required, see the box on page~\pageref{sidebar:statistically_optimum_beamforming}.
For the computation of the beamformer coefficients, \difftwo{the \gls{ATF} $\vect{a}_f$ and the inverse of $\scmtrue_{\vect{n},f}$ are required, see the box on page~??. The former can be obtained by computing the dominant eigenvector of  $\scmtrue_{\vect{x},f}$. In \mbox{\cite{zhang2021adl}}, it has been found that these two computations can be more reliably implemented with a \gls{RNN},}{the \glspl{SCM} of speech and noise are required, see Eq.~\eqref{eq:mvdr}.
In \cite{zhang2021adl}, it has been found that the \glspl{SCM} can be estimated more reliably directly with a \gls{RNN}},  in particular in the presence of time-varying statistics. \diffsimple{}{It has even been proposed to directly estimate the beamforming coefficients $\vect{w}_f$ by a neural network \cite{Xiao2016}.}

Subsequently, it was argued that the \gls{STFT} itself is a limitation, because of its uniform frequency resolution and its short-time stationarity assumption.
%and the conflicting requirements for a large window to adhere to the \gls{MTF}, and a small window to have high time resolution.
In \cite{gu2022towards}, the \gls{STFT} was replaced by learnable filters, and the beamforming operation was carried out in the time domain.
%The overall system, consisting of learnable encoder/decoder, mask estimator and beamformer weight computation was trained end-to-end.
The only remaining model-based component was the actual beamforming operation itself.

An even more radical step was taken in \cite{tesch2022insights}, where beamforming was abandoned completely, because the linear operation of beamforming was considered suboptimal: In the traditional model-based approach, noise reduction by beamforming is achieved by an \gls{MVDR} beamformer, followed by a spectral postfilter.
The first ensures a distortion-less response to the target signal, while the second improves the noise suppression.
For this model it has been shown that the \gls{MVDR} beamformer is a sufficient statistic in the Bayesian sense for estimating the clean speech signal $s$
%\cite{balan2002microphone, tesch2021nonlinear}, 
\cite{balan2002microphone}, if the noise has a Gaussian distribution: 
\begin{align*}
	p(s|\vect{y}) = p(s|\hat{x}^{\text{(MVDR)}})
\end{align*}
where $\hat{x}^{\text{(MVDR)}} = \vect{w}^{\text{(MVDR)}} \vect{y}$.
This important result says that linear filtering is optimal, because no information is lost for estimating $s$.

However, optimality is lost if the noise is no longer Gaussian.
For the model of \cref{eq:mtf}, Hendriks et al. have shown that the \gls{MMSE} optimal speech estimator in the presence of noise, which is modeled as a mixture of Gaussians, is a nonlinear, jointly spatial and spectral filter that cannot be decomposed into a cascade of spatial and spectral processing \cite{hendriks2009optimal}.
For speech enhancement tasks, non-Gaussianity is an important issue.
For example,  competing speakers are an example of non-Gaussian distortions.
Indeed, significant performance gains can be achieved with a nonlinear processor in the presence of non-Gaussian distortions. %\cite{tesch2021nonlinear}.
However, the nonlinear processor is very complex, and there is no realistic way to implement an enhancement algorithm based on this model.
It does not come to a surprise that the most straightforward choice of a non-linear processor is a \gls{DNN}.
With suitable \gls{DNN} architectures performance improvements have been achieved over beamforming \cite{tesch2022insights}.

However, it is difficult to compare the systems mentioned in this section across the different publications, because the reported results had been obtained on different data sets.

%\Cref{fig:model_data} illustrates the trend that ever more model-based components are absorbed in a \gls{NN}: From neural mask estimation \cite{Heymann_2015} to neural beamformer weight estimation \cite{xiao2016deep} to all-neural spatial filtering in the time-freuency domain \cite{li2021mimo} to finally  all-neural spatial filtering in the time-domain \cite{gu2022towards}.

%\begin{figure}
%	\begin{center}
%		\includegraphics[width=0.7\textwidth]{figures/model-based_vs_data-driven.png}
%		\caption{model-based vs data-driven trends: there is a trend to absorb more and more model-based system components in neural networks.
% Besseres Bild!}
%		\label{fig:model_data}
%	\end{center}
%\end{figure}

%%%%%%%%%%%%%%%%%%%%%%%%%%%%%%%%%%%%%%%%%%%%%%%%%%
\section{Conclusions}
\label{sec:conclusions}
%%%%%%%%%%%%%%%%%%%%%%%%%%%%%%%%%%%%%%%%%%%%%%%%%%%

In this contribution we have contrasted model-based, data-driven and hybrid approaches to speech enhancement, with a focus on multi-channel beamforming-inspired methods.
We illustrated the pros and cons of signal processing and neural approaches and found that blending the two often gives effective solutions that combine the best of both worlds.

From the literature study we observe a trend to replace model-based enhancement by hybrid and even by purely data-driven approaches.
In fact, even very established representations and algorithms are being replaced by learnable components.
Of course, one reason is that researchers want to explore the potential of new tools, which might be more rewarding than optimizing already very well-engineered techniques.
A more serious reason is, in our view, that speech enhancement has reached a degree of maturity that even minor approximations of models are no longer acceptable.
We illustrated this last point with the example of acoustic beamforming, where even established models, such as frequency-domain processing and linear beamforming, are challenged by neural approaches that, in principle, can overcome the simplifying assumptions of any model.
% In fact, we observe that the state of the art on several databases is held by purely neural architectures, see the footnote on page~\pageref{footnote_leaderboards}.

However, certain reservations have to be made.
First, the network is only as good as representative the training data is of the later field data where the network is going to be used.
Concerns remain about robustness: The network behavior can be arbitrarily poor in case of unexpected input, e.g., out of domain data \cite{zmolikova2023masked, VINCENT2017535}.
A model-based component may be inferior to a learnt module on the data sets, the learnt model is optimized for.
But the model can add to the robustness, since it introduces valid physical knowledge to the system, serving as a kind of ``regularizer'' to prevent the data-driven component from deviating too far from what is considered physically reasonable.
%This is in line with recent developments in physics-driven machine learning in the medical and imaging domain.\footnote{Special Issue of IEEE Signal Processing Magazine on Physics-Driven Machine Learning -- Applications in Computational Imaging, vol. 40, nos 1 and 2, 2023.} 
We have seen that hybrid approaches offer the possibility to combine the advantages of both worlds.
A good example is \gls{GSS} \cite{boeddeker2018chime5}, which, by combining data-driven and model-based parameter estimation, can boost the performance of a model-based enhancement approach while preserving the adaptation capability to the test data.
This may explain its robustness to highly challenging recording conditions such as the  CHiME challenge data\diffsimple{}{ \cite{Watanabe2020CHiME6}}.

There is a second reservation to be kept in mind: the network has to learn an ever more complex mapping function, which requires a larger network, in turn asking for more training data.
Thus, an important prerequisite of the trend towards learnable components is the availability of larger and more diverse \diffsimple{}{and realistic} training data.
A decision between a model-based component and its data-driven alternative should, therefore, be taken not only in terms of performance on a single data set but in terms of robustness and reliability on a wide range of diverse input data.

\bibliographystyle{IEEEtran}
\bibliography{IEEEfull,overview_refs}

\end{document}